\documentclass[12pt,aps,floatfix]{revtex4}
\usepackage{graphicx}
\usepackage{psfig}
\def\del{\partial}

\def\be{\begin{eqnarray}}
\def\ee{\end{eqnarray}}
\def\bea{\begin{array}}
\def\eea{\end{array}}
\def\bei{\begin{itemize}}
\def\eei{\end{itemize}}

\newcommand{\mat}{\left ( \begin{array}{cc}}
\newcommand{\emat}{\end{array} \right )}
\newcommand{\matf}{\left ( \begin{array}{cccc}}
\newcommand{\ematf}{\end{array} \right )}
\newcommand{\matt}{\left \begin{array}{ccc}}
\newcommand{\ematt}{\end{array} \right )}
\newcommand{\vect}{\left ( \begin{array}{c}}
\newcommand{\evect}{\end{array} \right )}

\newcommand{\nn}{\nonumber }

\newcommand{\beq}{\begin{eqnarray}}
\newcommand{\eeq}{\end{eqnarray}}

\begin{document}
\title{SUM RULES FOR THE DIRAC SPECTRUM OF THE SCHWINGER MODEL}

\author{L. Shifrin and J.J.M.  Verbaarschot}
\affiliation{Department of Physics and Astronomy, SUNY Stony Brook,\\
Stony Brook, NY 11794,\\
e-mail: leonid.shifrin@stonybrook.edu, 
jacobus.verbaarschot@stonybrook.edu} 

\begin{abstract}

The inverse eigenvalues of the Dirac operator in the Schwinger model satisfy
the same Leutwyler-Smilga sum rules as in the case of QCD with one flavor.
In this paper we give a microscopic derivation of these sum rules 
in the sector of arbitrary topological charge. 
We show that the sum rules can be obtained from the clustering property
of the scalar correlation functions.
This argument also holds
for other theories with a mass gap and broken chiral symmetry such as QCD
with one flavor. For QCD with several flavors a
modified clustering property is derived from the low energy chiral Lagrangian.
We also derive sum rules for a fixed external gauge field.
Both types of sum rules are obtained from the bosonized version 
of the Schwinger model.

 In the sector of topological charge $\nu$ the sum rules
are consistent with a shift of the Dirac spectrum away from zero by
$\nu/2$ average level spacings. This shift is also required to obtain
a nonzero chiral condensate in the massless limit.
Finally, we discuss the Dirac spectrum for a closely 
related two-dimensional theory with a gauge field action  that is
quadratic in the gauge field and has applications  
to the quantum Hall effect
and $d$-wave super-conductors.

\end{abstract}
\maketitle
\newpage
\tableofcontents
\newpage
\section{Introduction}

The Schwinger model \cite{Schwinger}, electrodynamics in $1+1$ dimension is 
one of the few quantum field theories that can be solved 
analytically \cite{Lowenstein:1971fc,Jayewardena:1988td,SW}. 
Since the model contains nonperturbative features that are also found in QCD, 
such as a nonzero chiral condensate and instantons, it has added significantly
to our understanding of gauge theories in general. 
However, most interesting quantum field theories cannot be solved analytically, and one
has to rely on approximate techniques to investigate them. 
For nonperturbative phenomena, two approaches which both have been tested in
the Schwinger model, have been very
successful: lattice simulations and chiral perturbation theory. In particular, 
a large number of lattice studies have benefited from the Schwinger model 
\cite{lat-kogut,lat-az,lat-dil,lat-neu,lat-gatt,lat-far,lat-shai,
lat-hoel,lat-kis,lat-biet,biet-dil,durr-prd71,lat-durr,lat-poul}. 
Among others, we mention the study of chiral fermions on
the lattice \cite{lat-neu,lat-far,lat-gatt,lat-biet}, the analysis
of the continuum limit for staggered fermions \cite{lat-durr}, and the
effect of quenching on the Dirac spectrum \cite{lat-durrsh,lat-poul}.

For theories with a mass gap and a nonzero chiral condensate the mass dependence
of the partition function can be obtained from a chiral Lagrangian. 
In particular,
in the domain where the Compton wavelength of the Goldstone modes is much
larger than the size of the box only the constant fields have to be included
in the partition function so  that its mass dependence simply follows from
the mass term of the chiral Lagrangian \cite{LS}. Based on this result, it
was shown in \cite{LS} that the Dirac spectrum has to satisfy consistency
relations in the form of sum rules for inverse eigenvalues. 
These relations are not sufficient to determine the Dirac spectrum. 
However,
it turned out that the complete
low-lying Dirac spectrum can be obtained from a modified chiral 
Lagrangian \cite{Vplb,OTV,DOTV}, 
or equivalently, from chiral random matrix theory \cite{SV,V}.
In the Schwinger model, and also for QCD with $N_f =1$, there are no low-lying
excitations and the only contribution to the low-energy limit of the partition
function is the vacuum energy. Starting from this result, it is possible to 
derive sum rules for the inverse Dirac eigenvalues of 
the Schwinger model \cite{LS,smvac}.
However, to obtain a deeper understanding of these sum rules in the
context of quantum field theory, it of is interest to derive them directly
from the microscopic field theory. This was achieved in \cite{smvac} for the
simplest sum rule in the sector of zero topological charge.

In this paper we give a microscopic derivation of the simplest sum rule
in the sector of arbitrary topological charge $\nu$.
This requires a detailed
understanding of scalar correlation functions in the Schwinger model which
were derived before both in the plane \cite{Manias:1989bu} and on the torus
\cite{SW,us,steele-VZ,steele,Azakov:2001pz}. It turns out that the clustering
property of the scalar correlation  
functions alone is sufficient to derive the (gauge field averaged) sum rules. 
Below we will give a derivation based on this property.
A third way to obtain Leutwyler-Smilga sum rules is to start 
from the bosonized form of the massive Schwinger model. 
In the microscopic derivation of sum rules we first derive 
a more informative quantity -- the same sum rules
for an arbitrary  fixed external gauge field (which was also obtained for
$\nu = 0$ in \cite{smvac}). The universal Leutwyler-Smilga sum rules then
follow after averaging over the gauge fields.

Analytical results obtained for the Schwinger model have been tested
elaborately by means of Monte-Carlo simulations. The numerical value
of the chiral condensate has been reproduced accurately 
\cite{lat-dil,lat-durr,lat-poul}. The index theorem was confirmed
and the fermionic would-be zero modes were identified \cite{lat-gatt,lat-far}.
Leutwyler-Smilga sum rules have been obtained both for zero and nonzero 
topological charge \cite{lat-far}. The distributions of the smallest 
eigenvalues of the Dirac operator for the Schwinger model 
\cite{lat-dil,lat-far,lat-durr,lat-poul} agree with analytical results 
obtained by means of chiral Lagrangians \cite{LS,smvac} and random 
matrix theory \cite{SV,V,indiv}. 

A theory that is closely related to the Schwinger model are random
Dirac fermions where the electromagnetic interaction is replaced
by a random gauge field interaction. Starting with the work of
Gade and Wegner \cite{gade} the behavior of the spectral density
near zero has attracted a great deal of attention in particular with
applications to the quantum Hall effect in mind \cite{ludwig,simons}.
An important difference with the Schwinger model is that in the condensed
matter literature the physically interesting case is the quenched model,
sometimes defined as the limit of zero flavors. The consensus is that
the density of states diverges for strong disorder as is the case for
the quenched Schwinger model \cite{doel,kenway,smvac,lat-poul}.
For weak disorder, on the other hand, the spectral density vanishes.
We will discuss the Dirac spectrum of this model and relate it to
the chiral condensate and the simplest Leutwyler-Smilga sum rule.

The organization of this paper is as follows. In an introductory chapter
we introduce the Leutwyler-Smilga sum rules and the Schwinger model and
discuss its main properties. A combinatorial derivation of sum rules
based on the 
clustering property of scalar correlators
is given in section \ref{combinatorial}. Section \ref{microscopic} contains 
a microscopic derivation of the simplest sum rule, and a derivation from
the bosonized massive Schwinger model is given in section \ref{bosonization}.
It is explained in section \ref{clustering} that modified sum rules
for two or more flavors can be related to modified clustering properties
of the scalar correlators.
A random matrix derivation of sum rules
is given in section \ref{random}. In this section we also study the
fluctuations of the fermion determinant. In the first part of 
section \ref{discussion}
we discuss the effect of topology on the Dirac spectrum and the
mechanism of the formation of a chiral condensate. In the second
part of section \ref{discussion} we discuss the behavior of the 
Dirac spectrum for random fermions.
Concluding remarks are made in section \ref{conclusions}.

\section{General Framework}
\subsection{Sum Rules}
In this paper we study the eigenvalues of the Dirac operator
by means of sum rules for its inverse eigenvalues introduced in
\cite{LS}. The eigenvalues of the anti-hermitian Dirac operator are
defined by
\be 
D\phi_k = i \lambda_k \phi_k.
\ee
In sector with
topological charge $\nu$ the Dirac operator has exactly $\nu$ zero
eigenvalues that are not paired. All other eigenvalues occurs
in pairs $\pm \lambda_k$.
In the sector of topological charge $\nu$ the
partition function is defined by
\be
Z^\nu(m) = m^\nu \int dA_\mu^\nu {\det}'[iD] e^{-S(A_\mu^\nu)},
\label{znu0}
\ee
where the integral is over gauge fields in the sector
of topological charge $\nu$ weighted by the gauge field action $S(A_\mu)$. 
As indicated by the prime, 
\be
{\det}'[iD] = \prod_{\lambda_n \ne 0} \lambda_n^2,
\ee
the zero eigenvalues are not included 
in the determinant resulting in the pre-factor $m^\nu$.
At fixed $\theta$ angle the partition function is given by
\be
Z(m,\theta) = \sum_{\nu = -\infty}^\infty e^{i\nu \theta} Z^\nu(m).
\label{ztheta}
\ee

Below we will use two types of averages of an operator $O$ denoted by single 
and double brackets.
The first type is defined by (notice that $Z^0(m=0)=Z(m=0,\theta)$)
\be
\langle {O} \rangle^{\nu} 
= \frac 1{Z^0(m=0)}\int DA_\mu^\nu O[A_\mu^\nu] {\det}'(iD) e^{-S(A_\mu^\nu)}, 
\label{aver-full}
\ee
and the second type by \cite{LS}
\be
\langle\langle {O} \rangle\rangle^{\nu} 
= \left . \frac {m^\nu} {Z^\nu(m)}\right |_{m=0}
\int D A_{\mu}^\nu O(A_\mu^\nu) {\det}'(iD)   e^{-S(A_\mu^\nu)}.
\label{aver-LS}
\ee

To obtain sum rules for the inverse Dirac eigenvalues 
in a sector of topological charge $\nu$ we expand the
fermion determinant 
in powers of $m$,
\beq
{\det}'\lbrack i\!\not\!\!D+m \rbrack
= {\det}' \lbrack i\!\not\!\!D \rbrack
\prod_{n\ne0}
\Big(1+m^2\sum_{n>0}\frac{1}{\lambda^{2}_n}
+ m^4 \sum_{n_i >0 \atop n_1\ne n_2} \frac 1{\lambda_{n_1}^2\lambda_{n_2}^2}
+\cdots \Big )
\label{determ_expand_lead_m}.
\eeq
For the small mass expansion of the partition function we thus obtain
\be
\frac{Z^{\nu}(m)}{Z^{0}(m=0)}&=&m^{\nu}\left\langle {\det}'[iD]\right
\rangle^\nu
+ m^{\nu+2}\left\langle \sum_{n>0}\frac{1}{\lambda^{2}_n}\right\rangle^{\nu}
+ m^{\nu+4}\left \langle \sum_{n_i >0 \atop n_1\ne n_2} 
\frac 1{\lambda_{n_1}^2\lambda_{n_2}^2}\right \rangle^\nu
+\cdots 
\nn \\
&=& m^{\nu}\left\langle {\det}'[iD]\right\rangle^\nu \Big[ 1 + m^2
\left\langle\left\langle 
 \sum_{n>0}\frac{1}{\lambda^{2}_n}\right\rangle\right\rangle^{\nu}
+ m^4 \left \langle\left\langle\sum_{n_i >0 \atop n_1\ne n_2} 
\frac 1{\lambda_{n_1}^2\lambda_{n_2}^2}\right \rangle\right \rangle^\nu
+\cdots \Big].
\label{smmass}
\ee 
where we have used the definition (\ref{aver-LS}) for the average
in the sector of topological charge $\nu$. Therefore, sum rules for
the inverse Dirac eigenvalues follow from the sub-leading terms in
the small mass expansion of the partition function in the sector
of topological charge $\nu$. If
\be
\frac{Z^{\nu}(m)}{Z^{0}(m=0)} = m^\nu(a_0 + a_2 m^2 + \cdots),
\ee
the simplest sum rule is given by
\be 
\left\langle\left\langle 
 \sum_{n>0}\frac{1}{\lambda^{2}_n}\right\rangle\right\rangle^{\nu}
= \frac{a_2}{a_0}.
\label{sum-def}
\ee
In \cite{LS} general arguments where given that, for a gauge theory
interacting with fermions according to the Dirac operator 
with a nonzero chiral condensate, this ratio is given by 
\be
\frac {\Sigma^2 V^2}{4(N_f+\nu)}.
\label{ls-2}
\ee
For $N_f =1$ this argument is particularly simple.
The $\theta$
dependence of the partition function, as can be observed from
(\ref{znu0}) and (\ref{ztheta}), is obtained from the replacement
$m \to m \exp(i\theta)$. Given that the vacuum energy is equal to
$-{\rm Re} (mV\Sigma)$, this results in the large volume partition function
\cite{LS}
\be
Z(m, \theta) = e^{mV \Sigma \cos\theta}.
\label{largeVpart}
\ee 
By inverting (\ref{ztheta}) we find that
the partition function in sector $\nu$ is given by \cite{LS}
\be
Z_\nu = I_\nu(mV\Sigma)= \frac 1{\nu! 2^\nu}(mV\Sigma)^\nu( 1+
\frac{(mV\Sigma)^2}{4(|\nu|+1)} +\cdots).
\label{zls}
\ee
This results immediately in the sum rule (\ref{ls-2}).
In this paper we will derive this sum rule by a microscopic calculation
in the Schwinger model.

\subsection{The Schwinger Model}

In this section we give a brief review of the Schwinger Model
\cite{Schwinger} which is massless QED in two dimensions. The Euclidean
Lagrangian is defined by
\be
\mathcal{L}=\frac 12 F_{01}^2  - 
\bar{\psi}\lbrack i\not \!\partial -g\not \!\! A\rbrack\psi,
\label{actionsw}
\ee
where the electric field strength is given by
\be
F_{01} = \del_0 A_1 - \del_1 A_0,
\ee
and
\be 
i\not \!\partial -g\not \!\! A = \gamma_\mu(i\del_\mu - g A_\mu),
\ee
with Euclidean gamma matrices defined by
\be
\gamma_0 = \sigma_1, \qquad \gamma_1 = \sigma_2,\qquad \gamma_5 = \sigma_3.
\ee
The Lagrangian of this model has a chiral symmetry which is broken by
the $U(1)$ axial anomaly. As a consequence the ``photon'' becomes massive
with mass given by $\mu = g/\sqrt \pi$. In this model, a local external
charge is screened by massless fermions. Therefore, the asymptotic
states contain no fermions, and the Schwinger boson is the only
physical particle. The theory is super-renormalizable with a  coupling 
constant $g$ (which has the  dimension of mass) that does not run.

The (massless) Schwinger model, being equivalent to a noninteracting gas of  
bosons, is exactly solvable and was solved many times, by different 
techniques (operator language, path integral, bosonization)
and on different manifolds \cite{Lowenstein:1971fc,Casher:1974vf,
Danilov:1980ez,Manton:1985jm,Hetrick:1988yg,Jayewardena:1988td,SW,
Joos:1990km,Roskies}. 
Also, many authors considered more specific 
features of the model,
such as certain correlation functions at zero and at finite temperature
\cite{bardakci,Rothe:1978hx,Manias:1989bu,us,steele-VZ,
steele,Grignani:1995cw,Durr:1996im,Azakov:2001pz}, 
the fermion determinant, zero modes, the index theorem and instantons 
\cite{Nielsen:1976hs,Nielsen:1977aw,Ansourian:1977qe,Patrascioiu:1979xj,
Hortacsu:1979fg,Hortacsu:1980kv,Seiler:1980yx,
SW,Fry:1992qz,Smilga:1993sn, Adam:1993fc},
the multi-flavor Schwinger model 
\cite{weis,shrock,Gatt-multi,shi-smil-multi,SmV,Hos-multi}.

The vector potential $A_{\mu}$ can be decomposed as
\be
A_{\mu}=-\epsilon_{\mu\nu}\partial^{\nu}\phi
+\partial_{\mu}\lambda ,
\label{vec_decomp}
\ee
with the last term being a pure gauge. The topological charge, $\nu$, 
of the
gauge fields is equal to the difference of the number of right handed and
left handed zero modes of the Dirac operator and is therefore
necessarily quantized. It is given by
\be
\nu &=& \frac g{4\pi} \int d^2 x \epsilon_{\mu\nu} F_{\mu\nu} = 
\frac g{2\pi} \int d^2x \del_\alpha^2 \phi.
\label{top_charge}
\ee 
By Stokes theorem the topological
charge is determined by the asymptotic behavior of $\phi$. 
One easily shows that the large distance asymptotics,
\be
\phi(x)\sim\frac{\nu}{2g}\log x^2 + {\rm const.} \ \ , 
\ \ \vert x\vert\rightarrow\infty , 
\label{bound_conds}
\ee   
results in a topological charge equal to $\nu$. 
The boundary conditions 
(\ref{bound_conds}) imply that the plane is compactified at infinity, for
example by stereographic projection to a sphere with radius 
$R$ \cite{Nielsen:1977aw,Ansourian:1977qe,Hortacsu:1979fg}.

Because of  a particular property of the 2d Dirac algebra,
\be
\gamma_{\mu}\gamma_5=-i\epsilon_{\mu\nu}\gamma_{\nu},
\ee
the Dirac operator can be written in   
in the form \cite{weis}
\be
i\not\!\! D_{\phi}=e^{g\phi\gamma_5}i\not\!\partial e^{g\phi\gamma_5} . 
\label{dirac_op}
\ee
Using this representation one easily finds the following explicit expressions
for the zero modes in the sector of topological charge $\nu$: 
\be
\psi_p(x)= \frac{1}{\sqrt{2\pi}}{(x^{+})}^p e^{-g\phi(x)} \vect 1 \\ 0 \evect
\ \ , \ \ p=0\cdots\nu-1 ,\qquad x^{\pm}=x_0\pm ix_1 .
\label{zero_modes}
\ee
Note that these zero modes are not normalized.
From the asymptotic behavior (\ref{bound_conds}), it is easy to see that 
for $p=0,\cdots, \nu -2$ the zero modes are normalizable. However, 
the zero mode for $p=\nu -1$ is not normalizable on the plane  (its norm 
diverges logarithmically). It has 
to be included nevertheless because it is normalizable on the compactified 
plane as required by the index theorem
\cite{Nielsen:1977aw,Ansourian:1977qe,Hortacsu:1979fg,bardakci,Jackiw:1984ji}. 
The pre-factor ${1}/{\sqrt{2\pi}}$ 
in the zero modes also follows from matching with 
compact manifolds \cite{Adam:1993fc}.

The unpaired zero modes explicitly break the 
$U_A(1)$ symmetry of the partition function.  
In the massless limit the chiral condensate 
$\langle\bar{\psi}\psi\rangle$ comes from the zero 
modes in the sectors $\nu=\pm 1$.
It  has a particularly simple form on the plane 
\cite{Lowenstein:1971fc,SW,smcon,Adam:1993fc},
\be
\Sigma\equiv \langle\bar{\psi}\psi\rangle=-\frac{\mu}{2\pi}e^{\gamma},
\label{cond}
\ee
where $\gamma$ is the Euler constant.
As a consequence of the Banks-Casher formula \cite{BC} the average
density of the low-lying Dirac eigenvalues is given by $\Sigma V/\pi$.
This also implies that if we take the thermodynamic limit before the
chiral limit, we find a nonzero value for the condensate from the
nonzero modes \cite{durr-prd71}. 
 
\subsection{Effective Action}

Using the action (\ref{actionsw}) and the decomposition (\ref{vec_decomp}),
the  partition function in the sector of topological charge $\nu$ 
is given by
\be
Z^{\nu}(m)&=&\int D\phi^{\nu} e^{-S(\phi^\nu)}
\int D\bar{\psi}D\psi e^{-\int d^2x \bar{\psi}
\lbrack i \ \not \!D+m\rbrack\psi} \nn \\
&=&
\int D\phi^{\nu} e^{-S (\phi^\nu)} \det[iD+m]\nn\\
&=&
\int D\phi^{\nu} e^{-S(\phi^\nu)} m^\nu \prod_{n>0} \lambda_n^2.
 \label{PI_part}
\ee
Here we have used the chiral symmetry of the nonzero Dirac eigenvalues. 
The infinite product over the eigenvalues has to be regularized
in the UV. This procedure is well known \cite{Fujikawa}, and results
in an anomalous contribution to the 
effective action. The result is
\cite{Hortacsu:1979fg,Roskies,SW}
\be  
\prod_{n>0}\lambda^{2}_n=\mathcal{C}\det { \cal N}\exp\Big(\frac{\mu^2}{2}\int 
d^2x\phi(x)\Delta\phi(x)\Big),
\ee
where $\mathcal{C}$ is an (infinite) normalization constant which 
drops out from all final results and ${\cal N}$ is the norm matrix of
the (unnormalized) zero modes,
\be
{\cal N}_{pq} = \int d^2 x \psi_p^\dagger(x) \psi_q(x).
\ee
The bosonic partition function in the sector of topological charge $\nu$ is
thus given by (to the lowest order in $m$)
\be
Z^{\nu}(m) = m^{\nu}\int D\phi^\nu \det\mathcal{N}e^{-\Gamma[\phi^\nu]},
\label{znu}
\ee
with the effective action $\Gamma$ equal to
\be
\Gamma(\phi) = \frac{1}{2}\int d^2 x \phi(x) 
[\Delta^2 -\mu^2 \Delta]\phi(x).
\label{effective}
\ee
This effective action $\Gamma$ defines the propagator
\be
\frac 1{p^4 + p^2 \mu^2} = \frac {1}{\mu^2}\bigg(\frac {1}{p^2} - 
\frac 1{p^2+\mu^2}\bigg).
\ee
In the coordinate space,
\be
\mathcal{G}(x)=-\frac{1}{4g^2}\log x^2-\frac{1}{2g^2}K_0(\mu\vert x\vert)
+ {\rm const.},
\label{greenx}
\ee
where $K_0(\mu\vert x\vert)$ is a modified Bessel function of the 
second kind, which is exponentially small at large distances.
The large distance behavior of this propagator is therefore determined 
by the first term which is a massless propagator
\be
\mathcal{G}(x)  = -\frac 1{4g^2}\log x^2 
+ {\rm const.}\,.
\label{asymg}
\ee 

The arbitrary constant in the  equation (\ref{greenx}) can only be obtained
after a suitable infrared regularization of the theory. 
For example, this constant can be determined 
by matching with the result on compact 
manifolds \cite{Smilga:1993sn,Adam:1993fc}. 

\subsection{Partition Function in the Sector of Topological Charge
$\nu$}
To evaluate  the partition function in the sector of topological charge 
$\nu$ (which we take positive for notational convenience; of course,
the partition function does not depend on the sign of the topological
charge),
we first rewrite the norm matrix. Using
the explicit expressions for the zero modes we find
\be
\det{\cal N} &=\frac{1}{(2\pi)^{\nu}}&\int d^2x_1\cdots d^2x_\nu 
\frac 1{\nu!} \sum_{\sigma \pi}
{\rm sg}(\sigma\pi) x_1^{*\, \sigma(0) }x_1^{ \pi(0) }\cdots
x_\nu^{* \sigma(\nu-1) }x_\nu^{ \pi(\nu-1) }
e^{-2g\sum_{q=1}^{\nu}\phi(x_q)}.\hspace{0.5cm}
\ee
The sums over permutations $\sigma$ and $\pi$ 
can be rewritten as the product of two 
determinants,
\beq
\det \mathcal{N}&=&\frac{1}{\nu!(2\pi)^\nu}\int d^2x_1\cdots d^2x_{\nu}
e^{-2g\sum_{q=1}^{\nu}\phi(x_q)}
\left|\begin{array}{cccc}1& x_1 & \cdots&x^{\nu-1}_1  \\
                         \vdots   &\vdots & &\vdots\\
                          1 & x_\nu & \cdots &x^{\nu}_{\nu-1}  
\end{array} \right |^2  = \nonumber \\ 
 &=& \frac{1}{\nu!(2\pi)^\nu}
\int d^2x_1\cdots d^2x_{\nu}e^{-2g\sum_{q=1}^{\nu}\phi(x_q)}
\prod_{i>j}^{\nu}\vert x_i-x_j\vert^2 ,  
\label{detN}
\eeq
where a well-known property of Vandermonde determinant was used. 
Using this  expression in (\ref{znu0}), we find for
the partition function to the lowest order in $m$,
\be
Z^{\nu}(m)=\frac{m^{\nu}}{\nu!(2\pi)^\nu}\int d^2x_1\cdots d^2x_{\nu}
\prod_{i>j}^{\nu}\vert x_i-x_j\vert^2\int D\phi e^{-\Gamma(\phi)}
 e^{-2g\sum_{q=1}^{\nu}\phi(x_q)}.
\label{partnu}
\ee
Since the effective action is Gaussian the path integral is simply given by
\be
\frac{1}{Z^0(m=0)}\int D\phi \ e^{-\Gamma(\phi)}  
e^{-2g\sum_{q=1}^{\nu}\phi(x_q)}=e^{2\nu g^2\mathcal{G}(0)+4g^2
\sum_{i>j}^{\nu}\mathcal{G}(x_i-x_j)}.
\ee
Using the explicit expression for $\mathcal{G}$ on the plane 
(\ref{greenx}) we observe that 
the factor \mbox{$\prod_{i>j}^{\nu}\vert x_i-x_j\vert^2$} is canceled 
by the asymptotic behavior of the Greens function. 
Thus, in the limit of large volume
\be 
\frac{Z^{\nu}(m)}{Z^{0}(m=0)}=\frac{m^{\nu}}{\nu!}
{\Big(\frac{e^{2g^2{\cal G}(0)} V}{2\pi}\Big)}^{\nu}
+O(m^{\nu+1}) .
 \label{part_fun_leading}
\ee
The chiral condensate can be expressed as
\be
\langle \bar \psi \psi \rangle = 
\frac 1{Z^0(m=0)}\frac 1V \lim_{m \to 0} \frac {Z^1(m) + Z^{-1}(m)}m.
\label{condef}
\ee
Since $Z^{-1}(m) = Z^1(m)$, the chiral condensate is given by
\be
\Sigma =  
\frac 1\pi e^{2g^2{\cal G}(0)},
\label{condens}
\ee
which can be used to eliminate ${\cal G}(0)$ from the partition 
function. Naively, eliminating ${\cal G}(0)$ in favor of 
the condensate is equivalent to fixing the constant in 
(\ref{greenx}). Notice that only the product of the square of
the normalization constant of the zero modes and $\exp(2g^2{\cal G}(0))$ is
fixed by this condition. 

\section{Combinatorial Derivation of Sum Rules}
\label{combinatorial}
In the sector of topological charge $\nu$ we find for the small
mass expansion of the  partition function
in a finite but large volume $V$
\be
\frac{Z^{\nu}(m)}{Z^{0}(m=0)}
=\sum_{n=0}^\infty \frac{m^{n}}{n!} 
\int d^dx_1 \cdots d^d x_n 
\langle \bar \psi \psi(x_1) \cdots\bar \psi \psi(x_n)\rangle^\nu.
\label{massex}
\ee
The average is over field configurations with topological charge
$\nu$ weighted by the gauge field action and the determinant of the Dirac
operator in the space of nonzero modes.
For one flavor, the terms in the 
expansion with $n< \nu$ vanish, since the number of  zero modes 
in the correlator is not sufficient to compensate the 
zeros of fermion determinant.
For $n =\nu$ we have
\be
\langle \bar \psi \psi(x_1) \cdots\bar \psi \psi(x_n)\rangle^\nu
=\langle \bar \psi_R \psi_L(x_1) \cdots\bar \psi_R \psi_L(x_n)\rangle^\nu.
\label{expnu}
\ee 
For massless quarks, this expectation value  is called a minimal correlator 
and saturated by zero modes \cite{bardakci,manias-cor,steele}.  If we use
the explicit expressions for the zero modes and take into account
their (anti-)symmetrization, we  recover (\ref{partnu}) and what followed.
However, since we are only interested in the leading large-volume behavior,
there is a simpler way - to use the clustering property: for all 
$|x_i-x_j|\to\infty$, 
\be
\langle \bar \psi_R \psi_L(x_1) \cdots\bar \psi_R \psi_L(x_n)\rangle^\nu 
\rightarrow\Big(\frac{\Sigma}{2}\Big)^{\nu},
\ee
so that, in agreement with (\ref{part_fun_leading}) \cite{Adam:1997wt}
\be
\frac{Z^{\nu}(m)}{Z^{0}\vert_{m=0}}=\frac{m^{\nu}}{\nu!}
{\Big(\frac{\Sigma V}{2}\Big)}^
{\nu}+O(m^{\nu+1}).  
\label{part_fun_a}
\ee
Let us consider sub-leading  orders in $m$ in  (\ref{massex}).
For $n>\nu$, the non-vanishing contributions have 
$n-\nu$ contractions from the nonzero mode 
part of the propagator. Since the massless propagator  
connects only states with the same chirality the expectation value
$\langle \bar \psi \psi(x_1) \cdots\bar \psi \psi(x_{\nu+1})\rangle^\nu =0$.
The first non-vanishing sub-leading term in (\ref{massex}) is 
\be
&&\frac{m^{\nu+2}}{(\nu+2)!}\int d^2x_1 \cdots d^2x_n d^2x d^2y 
\langle\bar{\psi}
\psi(x_1)\cdots\bar{\psi}\psi(x_n)\bar{\psi}\psi(x)\bar{\psi}\psi(y)
\rangle^{\nu}.
\label{part_fun_sec_order_m}
\ee
This correlator is sometimes called a non-minimal correlator
and was calculated in \cite{manias-cor,steele,Azakov:2001pz}. 
However, for our purpose 
we will only need the large distance limit of this correlator, which 
again follows from  the clustering property of the correlators.
By decomposing the field
into left handed and right handed components we find
\be
\int d^2x_1...d^2x_\nu d^2x d^2y 
\langle\bar{\psi}
\psi(x_1)\cdots\bar{\psi}\psi(x_\nu)\bar{\psi}\psi(x)\bar{\psi}\psi(y)
\rangle_{\nu}  \hspace*{5cm}\\
\hspace*{2cm}=(\nu+2)
\int d^2x_1\cdots d^2x_\nu d^2x d^2y 
\langle\bar{\psi}_L
\psi_R(x_1)\cdots\bar{\psi}_L\psi_R(x_\nu)
\bar{\psi}_L\psi_R(x)\bar{\psi}_R\psi_L(y)
\rangle^{\nu}.\nn
\ee
The combinatorial factor $\nu+2$ arises because the fermion bilinear
with opposite chiralities can be at each of the $\nu+2$ points. 
Using the clustering property for the expectation value we obtain
for the large volume limit of (\ref{massex})
\be
\frac{m^{\nu+2}}{(\nu+2)!}(\nu+2){\Big(\frac{\Sigma }
{2}\Big)}^{\nu+2}=\frac{m^{\nu+2}}{(\nu+1)!}
{\Big(\frac{\Sigma }{2}\Big)}^{\nu+2} . 
\label{combin_cancel}
\ee
The mass dependence of the partition function to this order is
thus given by
\be
\frac{Z^{\nu}(m)}{Z^{0}\vert_{m=0}}=\frac{m^{\nu}}{\nu!}
{\Big(\frac{\Sigma V}{2}\Big)}^{\nu}+
\frac{m^{\nu+2}}{(\nu+1)!}{\Big(\frac{\Sigma V}{2}\Big)}^{\nu+2}+O(m^{\nu+4}). 
 \label{part-clust}
\ee
Comparing this result with (\ref{smmass}), 
we arrive at the sum rule 
\be
\langle\langle\sum_{n\ne 0}\frac{1}{\lambda^{2}_n}\rangle\rangle^{\nu}
=\frac{1} {\vert\nu\vert+1}\frac{\Sigma^2 V^2}{2},
\label{sumnu}
\ee
which also incorporates the case of negative $\nu$. In \cite{LS} this result
was obtained from the mass dependence of the partition function at nonzero
vacuum angle. In our derivation we only used the clustering property
of expectation values. Therefore our derivation is also valid for QCD
with one massless flavor.
For the Schwinger model a derivation of the sum rule for zero topological
charge was given in \cite{smvac}.

The term of the order $m^{\nu+2l},\ \ l=2,3,_{\cdots}$ is not 
much more complicated. Again $\nu$ of the fermion bilinears are saturated
by zero modes. The remaining ones are contracted by the nonzero mode
Green's function, and therefore we have to choose $l$ of the
$\nu+2l$ fermion bilinears with opposite chirality   \cite{Adam:1997wt},
\be
&&\hspace*{-1cm}\int d^2x_1 \cdots d^2x_{\nu+2l} 
\langle\bar{\psi}
\psi(x_1) \cdots \bar{\psi}\psi(x_{\nu+2l})
\rangle^{\nu}  \\
&=&\left ( \begin{array}{c} \nu+2l \\ l \end{array} \right )
\int d^2x_1 \cdots d^2x_{\nu +2l}
\langle\bar{\psi}_L
\psi_R(x_1)\cdots \bar{\psi}_L\psi_R(x_\nu)
\bar{\psi}_L\psi_R(x_{\nu+1}\cdots\bar{\psi}_R\psi_L(x_{\nu+2l})
\rangle^{\nu}.\nn
\ee
Again using the clustering property we obtain for the large volume
limit of the $m^{\nu+2l}$ order term in (\ref{massex}),
\be
\frac{m^{\nu+2l}}{(\nu+2l)!}\frac{(\nu+2l)!}{(\nu+l)!l!}
{\Big(\frac{\Sigma V}{2}\Big)}^{\nu+2l}=
\frac{m^{\nu+2l}}{(\nu+l)!l!}{\Big(\frac{\Sigma V}{2}\Big)}^{\nu+2l}.
\ee
The term of order $m^{2l}$ in the expansion of 
(\ref{determ_expand_lead_m}) is given by
\be
m^{2l}{\det}'\lbrack i\!\not\!\!D \rbrack\sum_{{n_i>0}\atop 
{n_1\ne\cdots\ne n_l}}\frac{1}
{\lambda^{2}_{n_1}\cdots\lambda^{2}_{n_l}} \label{general_m_exp_term}.
\ee
Gathering all the pieces, we obtain the final result \cite{smvac,LS}
\be
\langle\langle\sum_{{n_i\ne 0\atop {n_1\ne\cdots\ne n_l }}}\frac{1}
{\lambda^{2}_{n_1}\cdots\lambda^{2}_{n_l}}\rangle\rangle^{\nu}
=\frac{\vert\nu\vert!}{2^l \ l! 
\ (\vert\nu\vert+l)!}\big (\Sigma V\big )^{2l}. 
 \label{gen_result}
\ee	
This sum rule is valid for any theory with the above clustering property.
In particular, it is valid for QCD with one flavor \cite{LS}.

\section{Microscopic Derivation of Sum Rules in the Schwinger Model}
\label{microscopic}
A microscopic derivation of the sum rule (\ref{sumnu}) 
in the sector of topological
charge $\nu =0$ was given in \cite{smvac}. In this section we 
generalize this derivation to arbitrary nonzero topological charge.
The main idea of the derivation may be summarized in the following 
identity
\be
\sum_{\lambda_n \ne 0}\frac 1{\lambda_n^2} = - {\rm Tr}\Big[G^{\nu\,2}\Big],
\label{sumg}
\ee
where $ G^\nu $ is the Green's function of the Dirac operator with
external field $A_\mu = - \epsilon_{\mu\rho} \del_\rho \phi$.
The equality (\ref{sumg}) follows from the spectral 
representation of the
Green's function 
\be
G^\nu(x,y) = \sum_{n\ne 0} \frac{\psi^{\phi}_n(x)\psi_n^{\phi\,^\dagger}(y)}
{i\lambda_n}  \label{standard_G},
\ee
and the fact that excited Dirac states are normalized to 1
(the trace is  both over Dirac indexes and spatial
coordinates). Therefore, we need the Green's function
$G^{\nu}(x,y)$ for an arbitrary background field $\phi$ in the sector of
topological charge $\nu$ (which is taken $\nu>0$ for convenience). 
Due to the index
theorem, the Dirac equation has exactly $\nu$ right-handed zero modes
for a generic field configuration in this class. 
The Green's function $G^{\nu}(x,y)$ is thus defined by  \cite{brown}
\be i\not\!\!
D_x G^{\nu}(x,y)=\delta(x-y)-\gamma_5^+P^\nu(x,y), \label{eqn_G}
\label{green-eq}
\ee 
where 
\be 
P^\nu(x,y) = \sum_{p=1}^{\nu}\frac{\psi_p(x)\psi_p^\dagger
(y)} {\int d^2z \psi_{p}^\dagger(z)\psi_p(z)},
 \label{eigen} 
\ee
 is the projector on the subspace of zero modes, 
$\psi_p(x),\,\, p=1,\cdots, \nu$,
and
\be
\gamma_5^+ =\frac 12(1+\gamma_5).
\ee
In other words, $G_{\nu}(x,y)$ 
is the Green's function in the space of the nonzero modes.
The explicit solution of (\ref{eqn_G}) is given by:
\be 
  G^{\nu}(x,y) = (1-\gamma_5^+P^\nu)\tilde G^{\nu}(1-\gamma_5^+P^\nu), 
\label{green-phi}
\ee 
where, because  $i\not\!\! D_{\phi}=e^{g\phi\gamma_5}i\not\!\partial
e^{g\phi\gamma_5}$,
\be
\tilde G^{\nu}(x,y)\equiv e^{-g\phi(x)\gamma_5}
G_0(x-y)e^{-g\phi(y)\gamma_5} \label{G_zero_nu}. 
\ee 
The free two-dimensional Dirac propagator  $G_0(x,y)$ is given by
\be
G_0(x,y) = \frac 1{2\pi}\frac{\gamma_\mu(x_\mu-y_\mu)}{(x-y)^2}.
\ee
In the representation $\psi_1(x) = e^{-g\phi(x)}$ and $\{\psi_k, k = 2, \cdots,
\nu\}$ is an orthogonal set perpendicular to $\psi$, one can easily
show that (\ref{green-phi}) is a solution of (\ref{green-eq}).
Using that  $\gamma_5^+ \tilde G^\nu \gamma_5^+ = 0$
we find the sum rule 
\be 
 \sum_{\lambda_n \ne
0}\frac 1{\lambda_n^2}
 &=& - {\rm Tr }\bigg[(1 -(1+\gamma_5)P^\nu)[G^{\nu}]^2\bigg]
\nn \\
 &=& - {\rm Tr} \bigg[(1+\gamma_5)(1 -
P^\nu)[G^{\nu}]^2 \bigg].
\ee 
In the second equality we have
used that in the term proportional to the identity, both helicities give
the same result. Using the explicit representation for the Green's
function and carrying out the trace over the $\gamma$ matrices, we find
\be 
 \sum_{\lambda_n \ne 0}\frac 1{\lambda_n^2} &=& 
- \int d^2 x d^2 y d^2 z (\delta(x-y)-P^\nu(x,y))tr\bigg[(1+\gamma_5)
e^{-g\phi(y)\gamma5} G_0(y,z) 
e^{-2g\phi(z)\gamma_5}G_0(z,x) e^{-g\phi(x)\gamma_5}\bigg]\nn \\ 
 &=& - \frac{2}{(2\pi)^2}\int d^2 x d^2 y d^2 z (\delta(x-y)-P^\nu(x,y))
e^{-g\phi(y) -g\phi(x)  +2g\phi(z)} \nn \\ &&\times
\frac
{(x-z)_\mu(z-y)_\mu - i\epsilon_{\mu\nu} (y-z)_\mu(z-x)_\nu} {(x-z)^2
(z-y)^2}\nn \\ 
&=& - \frac{1}{(2\pi)^2}\int d^2 x d^2 y d^2 z 
(\delta(x-y)-P^\nu(x,y))e^{-g\phi(y)
-g\phi(x)  +2g\phi(z)} \nn \\ 
&& \times
 \left[\frac {(x-y)^2 -2i\epsilon_{\mu\nu} (y-z)_\mu(z-x)_\nu}{(x-z)^2
(z-y)^2} -\frac 1{(x-z)^2} - \frac 1{(y-z)^2} \right ]. 
\ee  
In the last equality we have used that 
\be
(x-z)_{\mu}(z-y)_{\mu}=\frac{1}{2}\big( (x-y)^2-(x-z)^2-(z-y)^2\big).
\ee
The last two terms in the square brackets are independent of either
$y$ or $x$ and vanish after integration
as can be
seen from the representation (\ref{eigen}) 
\be 
\int d^2y P^\nu (x,y)
e^{-g\phi(y)} = e^{-g\phi(x)}. 
\ee 
Using that the integral over the $\delta$-function also vanishes, 
we find the sum rule 
\be 
\sum_{\lambda_n \ne 0}\frac 1{\lambda_n^2} 
&=&\frac{1}{(2\pi)^2}\int d^2 x d^2 y d^2 z
P^\nu(x,y)e^{-g\phi(y) -g\phi(x)  +2g\phi(z)} \frac
{(x-y)^2-2i\epsilon_{\mu\nu} (y-z)_\mu(z-x)_\nu} {(x-z)^2 (z-y)^2 }.\nn\\ 
\label{befaver}
\ee
 
The projector formula (\ref{eigen}) is written in terms of orthogonal modes.
However, since the zero modes (\ref{zero_modes}) are not orthogonal, we
require a projector formula that is valid for an arbitrary set of 
non-orthogonal modes. This is easily achieved by
using a  Lagrange interpolation 
such that the projector is equal unity for each of the zero modes. 
The required form is given by 
\be
P^\nu(x,y)= \frac 1{(\nu-1)!}\frac{1}{\det \mathcal{N}} \sum_{\sigma} 
{\rm sgn}(\sigma)
\left |\begin{array}{cccc} \psi^\dagger_1(y) \psi_{\sigma(1)}(x) &
(\psi_1,\psi_{\sigma(2)}) & \cdots& (\psi_1,\psi_{\sigma(\nu)})\\
\vdots & \vdots & & \vdots \\ \psi^\dagger_\nu(y) \psi_{\sigma(1)}(x) &
(\psi_\nu,\psi_{\sigma(2)}) & \cdots& (\psi_\nu,\psi_{\sigma(\nu)})\\
\end{array} \right |, 
\label{projector}
\ee 
where the normalization factor is the familiar determinant of the 
norm matrix, given by (\ref{detN}). The projector can be expressed 
in terms of Slater determinants as 
\be 
P^\nu(x,y)=\nu\frac{\int d^2x_2\cdots d^2x_{\nu}
\chi(x,x_2,\cdots,x_{\nu})\chi^\dagger(y,x_2,\cdots,x_{\nu}) } {\int
d^2x_1\cdots d^2x_\nu \vert\chi(x_1,\cdots,x_\nu)\vert^2} . 
\ee 
Here, $\chi(x_1,\cdots,x_\nu)$ is the Slater determinant of fermionic zero
modes. In this representation, it is particularly transparent that the projector
is not only independent of the choice of basis, but 
(unlike $\det \mathcal{N}$),
is also independent of the normalization of the zero modes, 
as of course it should be.
Using the explicit representation of the fermionic zero modes we
find the projector 
\be 
\nu\frac {\int dx_2 \cdots dx_{\nu}
\prod_{k=2}^{\nu} (x^+-x^+_k)(y^- -x_k^-) \Delta(\{x_2^+,\cdots,
x_{\nu}^+\})\Delta(\{x_2^-,\cdots, x_{\nu}^-\}) e^{-g(\phi(x)+\phi(y))-2g(
\phi(x_2) +\cdots+\phi(x_{\nu}))} } 
{\int dx_1 \cdots dx_{\nu}
\Delta(\{x_1^+,\cdots, x_{\nu}^+\})\Delta(\{x_1^-,\cdots, x_{\nu}^-\})
e^{-2g( \phi(x_1) +\cdots+\phi(x_{\nu}))} },\nn \\ 
\ee 
where $\Delta$ is a
Vandermonde determinant. Renaming $x$ by $x_1$ and $y$ by $x_{\nu+1}$,
the sum rule in the sector of topological charge $\nu$ can be written as
\be 
\sum_{\lambda_n \ne 0}\frac 1{\lambda_n^2}
 &=&\frac 2{(2\pi)^{\nu+2}(\nu-1)!}\frac{1}{\det \mathcal{N}} 
\int d^2 z \int dx_1 \cdots dx_{\nu+1}
e^{-2g(\phi(x_1)+\cdots+\phi(x_{\nu+1}))+2g\phi(z)}  
\nn\\ &&\times
\prod_{k=2}^{\nu} (x^+_1-x^+_k)(x^-_{\nu+1} -x_k^-) 
\Delta(\{x_2^+,\cdots x_{\nu}^+\})\Delta(\{x_2^-,\cdots x_{\nu}^-\})
\nn\\ &&\times 
  \frac {(x_1-x_{\nu+1})^2-
2i\epsilon_{\alpha\beta}(x_1-z)_\alpha(z-x_{\nu+1})_\beta} {(x_1-z)^2
(z-x_{\nu+1})^2 }.\nn\\ 
\ee 
Only the symmetric part of the expression in
the second line of this equation contributes to the integrand. Since
the expression is already symmetric in $x_2,...,x_{\nu}$, we only need
to symmetrize with respect to $x_1$ and $x_{\nu+1}$.
Then we get, 
\be &&{\cal S} \prod_{k=2}^{\nu}
(x^+_1-x^+_k)(x^-_{\nu+1} -x_k^-) \Delta(\{x_2^+,\cdots
x_{\nu}^+\})\Delta(\{x_2^-,\cdots x_{\nu}^-\})
  \nonumber \\ &&\times\frac
{(x_1-x_{\nu+1})^2/2-i\epsilon_{\mu\nu}(x_1-z)_\mu(z-x_{\nu+1})_\nu}
{(x_1-z)^2 (z-x_{\nu+1})^2 } \nn \\ &&=-\frac {\Delta(\{x_1^+,\cdots
x_{\nu+1}^+\}) \Delta(\{x_1^-,\cdots x_{\nu+1}^-\})}{\nu(\nu+1)}
\sum_{p\ne q} \frac
{(x_p-x_{q})^2/2-i\epsilon_{\mu\nu}(x_p-z)_\mu(z-x_{q})_\nu} {\prod_{l\ne
p} (x_p^--x_l^-) \prod_{l\ne q}(x_{q}^+-x_l^+)  (x_p-z)^2 (z-x_q)^2 } .\nn
\\ \label{step1} 
\ee 
The term with $p=q$ vanishes and can be included in
the sum. Using that 
\be 
\sum_p \frac 1{\prod_{k \ne p}(x_p-x_k)} = 0, 
\ee
we can replace the numerator 
\be
(x_p-x_{q})^2/2-i\epsilon_{\mu\nu}(x_p-z)_\mu(z-x_{q})_\nu 
\ee 
in (\ref{step1}) by 
\be 
&&(x_p-x_{q})^2/2 -(x_p-z)^2/2 -(x_q-z)^2/2
-i\epsilon_{\mu\nu}(x_p-z)_\mu(z-x_{q})_\nu \nn \\ &=&
-(x_p^+-z^+)(x_q^--z^-). 
\ee 
After simplifying the numerator we find for (\ref{step1})
\be
 \frac {2 \Delta(\{x_1^+,\cdots x_{\nu+1}^+\}) \Delta(\{x_1^-,\cdots
x_{\nu+1}^-\})}{\nu+1}
 \sum_{p,q}\frac 1 {\prod_{l\ne p} (x_p^--x_l^-) \prod_{l\ne
q}(x_{q}^+-x_l^+)  (x_p^--z^-) (x_q^+-z^+) }.\nn \\ 
\ee 
The sum in this
equation factorizes into the product of two sums. Each sum can be
expressed as a determinant 
\be
 \sum_{p}\frac {\Delta(\{x_1^-,\cdots x_{\nu+1}^-\})} {\prod_{l\ne p}
(x_p^--x_l^-)  (x_p^--z^-) } &=& \left | 
\begin{array}{ccccc} 
1/(x^-_1-z^-) & 1 & x_1^- & \cdots & x_1^{-\,\nu-1} \\ \vdots & & & & \vdots \\
1/(x^-_{\nu+1} -z^-) & 1 & x_{\nu-1}^{-} & \cdots & x_{\nu+1}^{-\, \nu-1}
\end{array} \right | \nn \\
&=& \frac {\Delta(\{x_1^-,\cdots
x_{\nu+1}^-\})}{\prod_p(x_p^- - z^-)}. 
\ee 
The determinant is easily
evaluated by multiplying each row by a factor $(x_k^--z^-)$. Exactly the
same simplification can be made for the $+$ variables. The sum rule
therefore simplifies to 
\be 
 \sum_{\lambda_n \ne 0}\frac
1{\lambda_n^2}
 &=&
 \frac 2{(2\pi)^{\nu+2}(\nu+1)!}\frac{1}{\det \mathcal{N}}
\int d^2 z \int dx_1 \cdots dx_{\nu+1}
e^{-2g(\phi(x_1)+\cdots+\phi(x_{\nu+1}))+2g\phi(z)} \nn\\ &&\times \frac
{\Delta(\{x_1^+,\cdots x_{\nu+1}^+\}) \Delta(\{x_1^-,\cdots
x_{\nu+1}^-\})} {\prod_{q=1}^{\nu+1}(z^+-x_q^+)(z^--x_q^-)}. 
\ee
This sum rule is valid for a arbitrary external gauge field 
$A_\mu = -\epsilon_{\mu\rho}\partial_{\rho}\phi$ in the sector
of topological charge $\nu$. The Leutwyler-Smilga sum rules (\ref{sum-def}) are 
obtained after averaging over the
fields $\phi(x)$ with the effective action $\det \mathcal{N}e^{-\Gamma[\phi]}$,
where $\Gamma[\phi]$ is defined in (\ref{effective}).
Notice that effective action  in the nontrivial topological sectors,
due to the presence of $\det \mathcal{N}$, 
is not only non-Gaussian in $\phi$ but also non-local in $\phi$.
However, the factor $\det \mathcal{N}$ cancels in the average of
the sum rule. The 
resulting path integral is Gaussian in $\phi$ and 
the average over the $\phi$ fields is given by a lowest order 
cumulant expansion. This results in
\be 
\left \langle \sum_{\lambda_n \ne 0}\frac
1{\lambda_n^2} \right \rangle
 &=&
 \frac 2{(2\pi)^{\nu+2}(\nu+1)!} \int d^2 z \int dx_1 \cdots dx_{\nu+1}
e^{2(\nu+2)g^2 G(0) +4g^2\sum_{k<l}^{\nu+1} G(x_k,x_l)
-4g^2 \sum_{k=1}^{\nu+1} G(x_k,z)}
\nn\\ &&\times \frac
{\Delta(\{x_1^+,\cdots x_{\nu+1}^+\}) \Delta(\{x_1^-,\cdots
x_{\nu+1}^-\})} {\prod_{q=1}^{\nu+1}(z^+-x_q^+)(z^--x_q^-)}.
\label{avers}
\ee
The second line of (\ref{avers}) is canceled by the
asymptotic behavior of the exponentiated Green's functions in the
first line of (\ref{avers}) (see (\ref{asymg})).
The large volume limit of the sum rule is therefore given by 
\be
\left \langle \sum_{\lambda_n \ne 0}\frac
1{\lambda_n^2} \right \rangle
 &=&
 \frac 2{(2\pi)^{\nu+2}(\nu+1)!} V^{\nu+2}e^{2(\nu+2)g^2 G(0)} =
 \frac {2}{(\nu+1)!}\bigg(\frac{\Sigma V}{2}\bigg)^{\nu+2},
\label{sumz}
\ee
where we used (\ref{condens}). Now, using (\ref{part_fun_a}), and the 
definition (\ref{aver-LS}), we finally obtain for the Leutwyler-Smilga
sum rule,
\be
\left \langle\left\langle \sum_{\lambda_n \ne 0}\frac
1{\lambda_n^2} \right \rangle\right \rangle_\nu
&=& \left . \frac {m^\nu Z^0(m)} {Z^\nu(m)}\right |_{m=0}
\left \langle \sum_{\lambda_n \ne 0}\frac
1{\lambda_n^2} \right \rangle
 \nn \\ &=& \left . \frac {m^\nu Z^0(m)} {Z^\nu(m)}\right |_{m=0}
\frac 2{(2\pi)^{\nu+2}(\nu+1)!} V^{\nu+2}
e^{2(\nu+2)g^2 G(0)}
= \frac 1{2(\nu+1)}\Sigma^2 V^{2}.\hspace{0.5cm}
\label{sumz2}
\ee
The result (\ref{sumz2}) is in agreement with the sum rule obtained 
by Leutwyler and Smilga \cite{LS} and with the result obtained from 
Random Matrix Theory \cite{SV}.

\section{Bosonization and Sum Rules}
\label{bosonization}
In this section we derive sum rules for the inverse Dirac eigenvalues 
starting from a bosonic description of the massive Schwinger Model.
In the particle physics literature, abelian bosonization goes back
to the work of Lowenstein and Swieca 
\cite{Lowenstein:1971fc}, who essentially bosonized massless Schwinger
Model, and  Coleman \cite{Coleman} and Mandelstam \cite{Mandelstam}.
The bosonized form of the massive Schwinger Model was studied
in  \cite{Coleman-Jackiw,Coleman-multi}. Most earlier works used an operator 
language. The path integral approach to bosonization, which we will use
below, was developed later, in 
many works, in particular in \cite{GamboaSaravi:1983xw,Naon:1984zp}.

It is well-known \cite{Coleman-Jackiw,Coleman-multi}
that the action of the massive Schwinger model can be written as
\be
S[\varphi]=
\int d^2x \Big[\frac{1}{2}\big(\partial_{\nu}\varphi\big)^2+
\frac{\mu^2}{2}{\varphi}^2-cm\cos(2\sqrt{\pi}\varphi-\theta)\Big]. 
\label{usual_bos_effaction}
\ee
with the constant $c={\mu}e^{\gamma_E}/(2\pi)\equiv |\Sigma|$. 
The value of this
constant is such that the chiral condensate is equal to its known
value (\ref{cond}).
At small $m$ this is a theory of a weakly self-coupled massive 
scalar field. The physics of this model is discussed in detail 
in \cite{Coleman-Jackiw,Coleman-multi}.
In the mean field limit, where $\phi =0$, we
obtain (\ref{largeVpart}) from the action (\ref{usual_bos_effaction}).
This directly leads to the Leutwyler-Smilga sum rules.

The first step in the microscopic derivation of sum rules is to 
eliminate the mass term of the $\varphi$ field by
\be
e^{-\int d^2 x \frac{\mu^2}{2}{\varphi}^2} = \int D(\Delta\phi) 
e^{-\int d^2x [\frac 12 (\Delta \phi)^2 + i\Delta \phi \varphi]} .
\label{intro-phi}
\ee

After partial integration of $(\del \varphi)^2$ 
and shifting $\varphi$ by $i\mu \phi$ we obtain 
the partition function
\be
Z(m,\theta) = \int D\varphi D\phi e^{-S_{\rm eff}[\varphi,\phi]},
\ee
with effective action given by
\be
S_{\rm eff}[\phi,\varphi]=\int d^2x \Big[\frac{1}{2}\phi\lbrack
\Delta^2-\mu^2\Delta\rbrack\phi + \frac{1}{2}\big(\partial_{\nu}
\varphi\big)^2 -\frac{1}{2}c m\Big[ e^{2g\phi+i \ 
2\sqrt\pi\varphi+i\theta} 
+e^{-2g\phi-i \ 2\sqrt\pi\varphi-i\theta}\Big]\Big].\nn \\ 
\label{intermed_bos_action}
\ee
The integral $D\phi$ is only over modes with $\Delta\phi$ 
not identically equal zero.
The partition function in the sector of topological charge $\nu$
is given by the Fourier transform of the partition function,
\be
Z^{\nu}(m)= \int_0^{2\pi}\frac{d\theta}{2\pi}
e^{i\theta\nu}Z(m,\theta).
\ee
The integral over $\theta$ is easily performed by expanding 
the integrand
in powers of $e^{i\theta}$,
\beq
Z^\nu(m)  &=& \int_0^{2\pi}\frac{d\theta}{2\pi}
e^{i\theta\nu}
\int D\varphi D\phi e^{-S_{\rm eff}(\varphi, \phi)}
\sum_{k,l=0}^{\infty} \frac{(c m/2)^{k+l}}{k!l!} 
\int d^2x_1 \cdots d^2x_k d^2y_1 \cdots 
d^2y_{l} \nonumber \\ 
&& \times 
\bigg [\prod_{q=1}^{k}e^{- 2g\phi(x_q)-  i \ 2\sqrt\pi \varphi(x_q)
-i\theta} \bigg]
\bigg[\prod_{p=1}^l e^{2g\phi(y_p) + 
i \ 2\sqrt\pi \varphi(y_p) +i\theta} \bigg].
\eeq 
It is clear that only the terms with $k = l+\nu$ contribute to the
integral. However, we can also perform the integral after 
shifting
\be
\phi \to \phi - \frac i{2g} \theta. \label{theta_shift}
\ee
In the effective action this results in the extra term
\be
i\theta \frac {g}{2\pi} \int d^2 x \Delta \phi.
\ee
The integral over $\theta$ therefore vanishes unless
\be
\frac {g}{2\pi} \int d^2 x \Delta \phi =\nu,
\ee
i.e. the topological charge of the fields $\phi$ is equal to $\nu$.

The path integral over $\varphi$ is Gaussian, 
and is given by
\be 
\int D\varphi 
\bigg[\prod_{p=1}^l e^{i \ 2\sqrt\pi \varphi(y_p)}\bigg] \bigg[
\prod_{q=1}^{\nu+l}e^{- i \ 2\sqrt\pi \varphi(x_q)}\bigg] 
e^{-S_0[\varphi]} =
\frac{\prod_{k<r}^l (y_k-y_r)^2\prod_{i<j}^{\nu+l}(x_i-x_j)^2}
{\prod_{q=1}^{\nu+l}\prod_{p=1}^l (y_p-x_q)^2}
\int D\varphi e^{-S_0[\varphi]} .\nn \\
\ee
with \footnote{It is assumed that an infrared regulator has been
introduced for the massless scalar field.} 
\be
S_0[\varphi]=\frac{1}{2}\int d^2x \big(\partial_{\mu}\varphi\big)^2.
\ee
Thus we end up with
\beq
             Z^{\nu}(m)&=&\int D\varphi e^{-S_0[\varphi]}
\int D\phi e^{-\Gamma[\phi]}\sum_{l=0}^{\infty} 
\frac{(c m/2)^{2l+\nu}}{(\nu+l)!l!} 
\int d^2y_1 \cdots  d^2y_l d^2x_1 \cdots d^2x_{\nu+l}
\nonumber \\ 
&&\times\frac{\prod_{k<r}^l (y_k-y_r)^2\prod_{i<j}^{\nu+l}(x_i-x_j)^2}
{\prod_{q=1}^{\nu+l}\prod_{p=1}^l (y_p-x_q)^2}
e^{2g[\phi(y_1)+\cdots+\phi(y_l)-\phi(x_1)-\cdots-\phi(x_{\nu+l})]}.
\label{exp1}
\eeq
The path integral over $\phi$ can be performed by a leading order
cumulant expansion. The Vandermonde determinants are again canceled
by the large distance asymptotic behavior of the Green's functions.
In the large volume limit we thus obtain 
\be
\frac{Z_\nu(m)}{Z_0(m=0)} = \sum_{l=0}^\infty
\frac{m^{\nu+2l}}{(\nu+l)!l!}{\Big(\frac{\Sigma V}{2}\Big)}^{\nu+2l},
\ee
which is equal to the result obtained from the clustering 
assumption
The sum rules are the same as derived in section
\ref{clustering}, 
\be
\langle\langle\sum_{{n_i\ne 0\atop {n_1\ne\cdots\ne n_l }}}\frac{1}
{\lambda^{2}_{n_1}\cdots\lambda^{2}_{n_l}}\rangle\rangle^{\nu}
=\frac{\vert\nu\vert!}{2^l \ l! 
\ (\vert\nu\vert+l)!}\big (\Sigma V\big )^{2l} .
\ee	
From this result one easily derives that 
$Z(m,\theta)= \exp(mV\Sigma\cos\theta) $ (see also \cite{LS,Adam:1997wt}).

We end this section with a conjecture for sum rules in a fixed external
$\phi-$field.
By now we realize that the field $\phi$ introduced in (\ref{intro-phi}) 
is indeed the same as
in previous sections. We can extract the sum rules from comparison with 
the partition function in eigenvalue representation. 
To that end, we rewrite the partition function as
\be
Z^{\nu}(m) &=& m^{\nu}\int D\phi e^{-S[\phi]} 
{\det}'\lbrack i\!\not\!\!D \rbrack \frac{{\det}' \lbrack i\!\not\!
\!D+m \rbrack}{{\det}' \lbrack i\!\not\!\!D \rbrack } \nn \\
&=&
\int D\phi e^{-\Gamma[\phi]}m^{\nu}\det \mathcal{N}^{\nu}\sum_{l=0}^{\infty} 
m^{2l} \sum_{{n_i\ne 0\atop {n_1\ne\cdots\ne n_l }}}\frac{1}
{\lambda^{2}_{n_1}\cdots\lambda^{2}_{n_l}}, 
\label{exp2}
\ee
where the formulas (\ref{determ_expand_lead_m}) and (\ref{general_m_exp_term}) 
have been used.
We remind that $\det\mathcal{N}^{\nu}$ is determinant of the norm matrix 
(\ref{detN}).  We conjecture that the integrands
of (\ref{exp1}) and (\ref{exp2}) are equal:
\beq
\sum_{{n_i\ne 0\atop {n_1\ne\cdots\ne n_l }}}\frac{1}{\lambda^{2}_{n_1}
\cdots\lambda^{2}_{n_l}} &=& \frac{\mathcal{C}}{(\nu+l)!l!\det\mathcal{N}^\nu}
\int  d^2y_1\cdots d^2y_l d^2x_1 \cdots d^2x_{\nu+l}  \\ 
&&\times \frac{\prod_{k<r}^l (y_k-y_r)^2\prod_{i<j}
^{\nu+l}(x_i-x_j)^2}{\prod_{q=1}^{\nu+l}
\prod_{p=1}^l (y_p-x_q)^2}e^{2g[\phi(y_1)+ \cdots
+\phi(y_l)-\phi(x_1)-\cdots-\phi(x_{\nu+1})]}.\nn
\eeq
For $l=1$ this conjecture was shown to be valid in 
 in section \ref{microscopic}.
 
\section{Clustering for Two of More Flavors in ChPT}
\label{clustering}
In multi-flavor QCD, chiral symmetry is broken spontaneously, with 
the appearance of massless Goldstone bosons. This results in long range
correlations that will modify the clustering property of scalar correlators
and will affect the sum rules for the inverse Dirac eigenvalues.
Starting from a chiral Lagrangian we will propose a modified clustering
relation. 

Let us first show that factorization leads to an incorrect result. 
To leading order in the quark masses the partition function in the 
sector of topological charge $\nu$ is given by
\beq
\frac{Z^{\nu}(m_1,\cdots,m_{N_f})}{Z^0(m_i =0)}
 =   \frac{(m_1 ... m_{N_f})^{\nu}}{[\nu !]^{N_f}}\int \prod_{i,j}dx_i^{(j)}
 \Big\langle\bar{\psi}\psi(x_1^{(1)})\cdots\bar{\psi}\psi(x_{\nu}^{(1)})
\cdots \bar{\psi}\psi(x_1^{(N_f)})\cdots\bar{\psi}\psi(x_{\nu}^{(N_f)}) 
\Big\rangle^{\nu}.\nn \\
\label{part_fun}
\eeq
If we assume clustering as in the case of one flavor we obtain 
\be
\Big\langle\bar{\psi}\psi(x_1^{(1)})\cdots \bar{\psi}\psi(x_{\nu}^{(1)}) 
\cdots \bar{\psi}\psi(x_1^{(N_f)})\cdots \bar{\psi}\psi(x_{\nu}^{(N_f)}) 
\Big\rangle^{\nu} \rightarrow 
{\Big(\frac{\Sigma}{2}\Big)}^{\nu N_f}, 
\ee
resulting in the partition function
\be
\frac{Z^{\nu}(m_1,\cdots,m_{N_f})}{Z^0(m_i =0)}
=  \frac{1}{[\nu !]^{N_f}} \ \bigg[ \frac{m_1\Sigma V}{2}
\bigg]^{\nu} \cdots\bigg[\frac{m_{N_f}\Sigma V}{2}\bigg]^{\nu}.
\label{wrong_partition}
\ee
However, instead of $1/[\nu !]^{N_f}$, the correct pre-factor 
\cite{LS} is given by
$\prod_{k=0}^{N_f-1}{k!}/{( \nu  +k)!}$.

The reason for the incorrect pre-factor in  (\ref{wrong_partition}) is 
that we did not account for the  spontaneous symmetry breaking of
$SU_A(N_f)$, which results in massless Goldstone bosons and long
range correlations.
As a consequence, the clustering property
does not hold directly even for the correct ground state 
(i.e. at fixed $\theta$). It is possible, as we will show below,
to  write down a modified clustering property. 

The clustering property follows from the large distance behavior of
the correlators. This is determined by the zero momentum term in the
chiral Lagrangian, i.e. by 
\be
Z_{\rm Low}(M,\theta) = \mathcal{N}\int\mathcal{D}U e^{V \Sigma{\rm Re} 
{\rm Tr}(MU^\dagger e^{i\theta/N_f}) },
\label{zlow}
\ee
with normalization constant ${\cal N}$ such that the partition function
is equal to unity for $M=0$.
For a diagonal mass matrix we obtain the expansion
\be
Z_{\rm Low}(m_f,\theta) = \sum_{k_1, \cdots ,k_{N_f}=0}^{\infty}
\int_{U \in SU(N_f)} dU 
\prod_{l=1}^{N_f }\bigg[\bigg(\frac{m_l V \Sigma}{2}\bigg)^{k_l}
\frac{(U_{ll}^\dagger e^{i\theta/{N_f}} + 
  U_{ll}e^{-i\theta/{N_f}}  )^{k_l}}{k_l!}\bigg] ,
\ee
which should be compared to the large volume limit of the QCD partition
function given by
\be
Z^{\rm QCD}(m_f,\theta) = \sum_{k_1, \cdots ,k_{N_f}=0}^{\infty}
\prod_{l=1}^{N_f }\frac{\big( m_l V \big)^{k_l} }{k_l!}
\langle \prod_{l=1}^{N_f}(\bar \psi_l \psi_l)^k_l\rangle_\theta .
\ee
We thus obtain the clustering relation
\be
\langle \prod_{l=1}^{N_f}(\bar \psi_l \psi_l)^{k_l}\rangle_\theta 
=
\int_{U \in SU(N_f)} dU 
 \prod_{l=1}^{N_f }\frac \Sigma 2(U_{ll}^\dagger e^{i\theta/{N_f}} +
  U_{ll}e^{-i\theta/{N_f}} )^{k_l}.
\label{cluster-u0}
\ee
If we isolate the phase $\xi_i$ of the diagonal matrix element $U_{ii}$ 
this can be rewritten more suggestively as
 \be
\langle \prod_{l=1}^{N_f}(\bar \psi_l \psi_l)^{k_l}\rangle_\theta 
=
\int_{U \in SU(N_f)} dU 
 \prod_{l=1}^{N_f }|U_{ll}|
\langle \bar \psi \psi \rangle_{(\theta/N_f) -\xi_l}^{k_l}, 
\label{cluster-u}
\ee
with $\langle \bar \psi \psi \rangle_{(\theta/N_f) -\xi_l}$ the
chiral condensate for one flavor and vacuum angle equal to
$(\theta/N_f) -\xi_l$. 
Notice that only the diagonal matrix elements of $U$ enter in this 
expression.
Since $|U_{ll}|\le 1$ we could 
interpret the factors $|U_{ll}|$ as the projection of  the 
one flavor condensates onto 
certain directions in  group space.

In general, the integral in (\ref{cluster-u0}) cannot be evaluated 
analytically.  
However, the situation simplifies enormously for the minimal correlators 
in the sector of fixed 
topological charge $\nu$. In this case we have  
$n_1=n_2=\cdots=n_{N_f}=\nu$ and the integral over $\theta$ is only nonzero
if no factors $U^\dagger_{ll}$ occur. This results in
\be
\langle \prod_{l=1}^{N_f}(\bar \psi_l \psi_l)^\nu\rangle^\nu 
=\left( \frac \Sigma 2\right )^{\nu N_f}
\int_{U \in SU(N_f)} dU 
 \prod_{l=1}^{N_f } U_{ll}^\nu.
\label{cluster-min}
\ee
This integral can be evaluated analytically (see appendix A) 
\be
\int dU  U_{11}^\nu U_{22}^n\cdots U_{N_f N_f}^\nu = 
\Big[\nu!\Big]^{N_f}\prod_{i=0}^{N_f-1}\frac{i!}
{(i+\nu)!} .
\label{udiagint}
\ee
The leading mass dependence of the partition function is obtained 
by  substituting the result (\ref{cluster-min}) 
into equation (\ref{part_fun}),
\be
Z^{\nu}(m_1,\cdots,m_{N_f}) 
=   \bigg[ \frac{m_1\Sigma V}{2}\bigg]^{\nu} \cdots 
\bigg[\frac{m_{N_f}\Sigma V}{2}\bigg]^{\nu}
\prod_{i=0}^{N_f-1}\frac{i!}{(i+\nu)!}.
\ee
This explains the suppression of the scalar correlator for $N_f \ge 2$.

\section{Relations with Random Matrix Theory}
\label{random}

The chiral random matrix model in the sector of topological charge $\nu$
and $N_f$ flavors with equal quark mass $m$ is defined by \cite{SV,V}
\be
Z^\nu_{N_f}(m) = C_n^{(1)}\int dW {\det}^{N_f}( D + m) 
e^{- n\Sigma^2 {\rm Tr} W W^\dagger},
\label{zrmt}
\ee 
where 
\be
 D
 =\left ( \begin{array}{cc} 0 &i W\\
                                   i W^\dagger  &0
\end{array} \right ),
\label{DRMTT}
\ee
and $W$ is a complex $n \times(n+\nu)$ matrix.
The eigenvalues of this model are distributed according to a semicircle.
The largest eigenvalue is equal to $1/\Sigma$. We use this eigenvalue 
to normalize the partition function such that it becomes dimensionless.
If we also normalize the partition function to unity for $N_f =0$ we
obtain the normalization constant
\be
C_n^{(1)} = e^{nN_f}\left (\frac n\pi  \right )^{n(n+\nu)} 
\Sigma^{2n(n+\nu) }.
\ee
We also have included the factor $e^{nN_f}$ which eliminates the constant
vacuum energy for $N_f \ne 0$.
We can evaluate the partition function (\ref{zrmt}) in the limit $m\to 0$
in two different ways. First,
using an eigenvalue representation of $W$, and second, using a $\sigma$-model
representation of the large $n$ limit of the 
partition function. Since the $\sigma$-model 
is exactly the partition function
(\ref{zlow}), we only discuss the first
approach in the next subsection. 

         \subsection{Eigenvalue Representation}

To obtain an eigenvalue representation of the partition function we use
the polar decomposition
\be
W = U \Lambda V^{-1},
\ee
where $ U \in U(n+\nu)/U^n(1)\times U(\nu)$ and $V\in U(n)$. The Jacobian of
this transformation is given by
\be
J =  \Delta(\{\lambda_k^2\}) 2^n\prod_{k=1}^n(\lambda_k)^{2\nu+1}.
\ee
The eigenvalues of $\Lambda $ are positive or zero, and  the integration
is over an ordered sequence of eigenvalues which will
be accounted for by a factor $1/n!$ below. To lowest non-vanishing order
in $m$, the eigenvalue representation of the partition function is 
thus given by
\be
Z^\nu_{N_f}(m) = C_n^{(1)}C_n^{(2)}  m^{\nu N_f} \frac 1{n!}
\int \prod_{k=1}^n 
2d\lambda_k \lambda_k^{2(\nu+N_f)+1} \Delta^2(\{\lambda_k^2\}) 
e^{-n\Sigma^2\sum_k\lambda_k^2},
\ee
where we have introduced a second constant,
\be
C_n^{(2)}= 2^{-n^2-n\nu}\frac{ {\rm vol}\,(U(n+\nu))
{\rm vol}\,(U(n))  }
{{\rm vol}\,(U(\nu)){\rm vol}^n\,(U(1))},
\ee
and
the volume of $U(n)$ is given by
\be
{\rm  vol \,}(U(n)) = \frac {(2\pi)^{n(n+1)/2}}{\prod_{k=0}^{n-1} k!}.
\ee
We rewrite the partition function in terms of the integration variables
\be
x_k = 2n \Sigma^2 \lambda_k^2.  
\ee
This results in
\be
Z^\nu_{N_f}(m) = C_n^{(3)}C_n^{(2)}  m^{\nu N_f} \frac 1{n!}
\int \prod_{k=1}^n 
dx_k x_k^{\nu+N_f} \Delta^2(\{x_k\}) 
e^{-x_k/2},
\ee
where
 \be
C_n^{(3)} = (2n)^{-[n(n+\nu) + n N_f]} \pi^{-n(n+\nu)} e^{nN_f}.
\ee
The integral over the $x_k$ was calculated by Forrester \cite{forrester-book}.
Our final result for the partition function is thus given by
\be
Z^\nu_{N_f}(m) = m^{\nu N_f} \Sigma^{-2n N_f}
C_n^{(3)}C_n^{(2)}  2^{n(\nu+N_f+n)} 
\prod_{j=0}^{n-1} j!(\nu+N_f+j)!.
\ee
Collecting all factors we find the partition function
\be
Z^\nu_{N_f}(m) = m^{\nu N_f} \Sigma^{-2n N_f}n^{-nN_f} e^{nN_f}
\prod_{j=0}^{n-1} \frac {( \nu+N_f+j)!}{(\nu+j)!}.
\ee
Using the Stirling formula, we find in the thermodynamic limit
\be
Z^\nu_{N_f}(m) = m^{\nu N_f} \Sigma^{-2n N_f}
\frac{ (2\pi)^{N_f/2} n^{N_f(N_f+2\nu)/2}} 
{\prod_{k=0}^{N_f-1}(\nu+k)!}.
\label{zrmtfin}
\ee

For $N_f =1$ we find the ratio
\be
\frac{Z^{\nu=1}_{N_f=1}(m)}{Z^{\nu=0}_{N_f=1}(m)} = m (n+1).
\label{zdemo}
\ee
According to (\ref{condef}) the condensate is given by
\be
\lim_{V\to\infty}\lim_{m\to 0} \frac 1{m V}\left [ 
\frac{Z^{\nu=1}_{N_f=1}(m)}{Z^{\nu=0}_{N_f=1}(m)}+
\frac{Z^{\nu=-1}_{N_f=1}(m)}{Z^{\nu=0}_{N_f=1}(m)}\right ].
\ee
In random matrix theory the volume is identified as $V =2n$ such that 
(\ref{zdemo}) results in a
condensate equal to unity. The reason for this incorrect result is
that we have evaluated the ratio of two products with the same number
of nonzero eigenvalues. To obtain a result with the correct 
dimensionality 
we have to regularize the determinants by only including eigenvalues
below a given energy \footnote{For overlap fermions \cite{neuberger} 
achieved by pairing the zero eigenvalues with the largest eigenvalues. 
In our case, the largest eigenvalue is equal to $1/\Sigma$. 
If we extend the
Dirac operator for $n-1$ modes and $\nu =1$ with 
this eigenvalue, we find the ratio
(for $N_f =1$)
$
{\left . Z^{\nu=1}_{N_f=1}(m)\right|_{n \to n-1}}/
{\left . Z^{\nu=0}_{N_f=1}(m)\right |_{n}} 
= m n \Sigma,
$
where we have used (\ref{zrmtfin}). 
This result also explains that in the case of overlap fermions
the correct chiral condensate is obtained in the Schwinger model 
if the limit $m\to 0$ is taken
first \cite{lat-durr}. }.
For $\nu =1$ we have on average one eigenvalue
less with absolute value below a given energy than for $\nu =0$.
This results in an overall
factor of $1/p\Delta$, where $p$ is the number of eigenvalues below
the cut-off energy and $\Delta$ is the spacing between the eigenvalues. 
Instead of using a fixed energy cut-off, we calculate the ratio of
the partition functions for a fixed total number of eigenvalues meaning that
in the sector of topological charge $\nu$ we make the
replacement $n \to n - \nu/2$ in (\ref{zrmtfin}). We finally find the
ratio
\be 
\frac{Z^\nu_{N_f}(m)}{Z^0_{N_f}(m=0)} = (mn\Sigma)^{\nu N_f}
\prod_{k=0}^{N_f-1}\frac {k!}{(k+\nu)!} ,\nn \\
\label{anom}
\ee
in agreement with general arguments given in \cite{LS}. 
This result also follows from the fact
that the large $N$ limit of the random matrix model is given by
the nonlinear sigma model (\ref{zlow}) \cite{SV}. 
As was already found in \cite{SV}, the result
(\ref{anom}) confirms that the chiral anomaly is contained
in random matrix theory.

Since the finite volume partition function for $N_f=1$ is a Bessel
function, it is tempting to identify its zeros with the average
position of the eigenvalues. In the random matrix model it can
be easily checked whether this identification is justified. 
Numerically, it turns out that the ratio obtained from the 
average eigenvalues is a factor $2/\pi$ smaller than the result
given in (\ref{zdemo}). A second question we have asked is 
whether the average determinant is  mainly given by the 
product of the average eigenvalues or whether it is due to 
the fluctuations of the eigenvalues. It turns out that the product
of the average eigenvalues (with the determinant included in the
weight) is an order of magnitude larger than the average determinant.

\section{Discussion}
\label{discussion}
\subsection{Spectral Duality}

The chiral condensate can be obtained in two ways. First, by taking
the chiral limit after the thermodynamic limit,
\be
\Sigma^{(1)} =  \lim_{m\to 0} \lim_{V\to \infty} \frac 1V \left \langle
\frac 1{i\lambda_k +m} \right \rangle,
\ee
and, second, by inverting the two limits,
\be
\Sigma^{(2)} =   
\lim_{V\to \infty} \lim_{m\to 0} \frac 1V \left \langle
\frac 1{i\lambda_k +m} \right \rangle.
\ee 
In the first case, the condensate arises as a consequence of 
spontaneous symmetry breaking, and in the second case it is due to
the anomaly or instantons. The two condensates are not 
necessarily equal. For example, if we restrict the partition
function to gauge field configurations with zero topological
charge, the second definition gives zero. By spectral duality we
mean that
the two condensates are equal for the $\theta$ - vacuum
\cite{sum-poul}. In the second case, when
the contribution to the condensate comes from the sector with
topological charge equal to one, the eigenvalues for $\nu = 1$ are
shifted
such that the condensate is equal to the one obtained from the
first definition. The first definition implies the Banks-Casher
formula \cite{BC}
\be 
\Sigma^{(1)} = \frac {\pi \rho_0}V.
\ee
This implies that the average spacing of the eigenvalues is
given by 
\be
\Delta = \frac 1{\rho(0)} = \frac \pi{\Sigma V}.
\ee
If we assume that the position of the eigenvalue $\lambda_n$ is 
given by $n\Delta$, we obtain the sum rule
\be
\sum_{n >0} \frac 1{n^2\Delta^2} = \frac 1 6{\Sigma^2 V^2},
\ee
which explains the functional dependence of the sum rule. 

As we will show next, the large $\nu$ limit of the sum rule can 
be obtained from the
large $\nu$ limit of the average position of the eigenvalues. 
In the
sector of topological charge $\nu$ the ``average'' position of the eigenvalues
can be expressed in terms of 
the zeros, $j_{\nu,k}$, of the Bessel functions \cite{LS}, 
\be
\lambda_k = \pm \frac {j_{\nu,k}}{\Sigma V}.
\label{128}
\ee
For large $k$ at fixed $\nu$ one one finds
\be
\lambda_k 
\sim j_{\nu,k}^{\rm as} \Delta 
\sim (k + \frac \nu 2 -\delta)\Delta \qquad {\rm with}
\qquad \delta = \frac 14 .
\label{lamas} 
\ee
Here, $\Delta\equiv 1/\rho(0)$ is the average spacing of the eigenvalues.
This formula does not accurately give the position of the eigenvalues for
large $\nu$ and finite  $k$. In this case one can use a
uniform asymptotic expansion of the zeros given by
\be
j_{\nu, k} \sim \nu z_k,
\ee
with $z_k$ implicitly defined by
$
(4k-1)\pi = 4\nu(\sqrt{z_k^2-1} - \arccos(1/z_k)).
$
This results in the correct large $\nu$ limit of the sum rule
\be
\sum_k \frac 1{\lambda_k^2} &\sim& \Sigma^2 V^2
\int_{0}^\infty dk \frac 1{j_{\nu,k}^2}
= \frac{\Sigma^2 V^2 }{\pi \nu} \int_1^\infty dz\frac{\sqrt{z^2-1}}{z^3}
=\frac {\Sigma^2 V^2}{4\nu}.
\ee

Let us now
calculate the condensate according to the second definition 
\be
\Sigma^{(2)} 
= \frac 1V \frac{\langle\prod_k'(i\lambda_k+m)\rangle_{\nu=1}}
         {\langle \prod_k (i\lambda_k+m)\rangle_{\nu=0}}
+\frac 1V \frac{\langle\prod_k'(i\lambda_k+m)\rangle_{\nu=-1}}
                  {\langle \prod_k (i\lambda_k+m)\rangle_{\nu=0}}.
\label{siganom}
\ee
Using the expression (\ref{zls}) for the partition function, this
can be rewritten as
\be
\Sigma^{(2)} = \lim_{m\to 0} \frac 1V 
\left [ \frac{I_1(mV\Sigma)/m}{I_0(mV\Sigma)}+
\frac{I_{-1}(mV\Sigma)/m}{I_0(mV\Sigma)}\right ] = \Sigma.
\ee
The ratio in (\ref{siganom}) can be
interpreted as the ratio of the 
expectation values
of the fermion determinant in the topological sector $\nu =1$ and
$\nu =0$. 

We will calculate the averages in (\ref{siganom}) by replacing the
eigenvalues according to (\ref{128})
\be
\lim_{m\to 0}\frac 1V \frac{\langle\prod_k'(i\lambda_k+m)\rangle_{\nu=1}}
         {\langle \prod_k (i\lambda_k+m)\rangle_{\nu=0}}
&\approx&
\lim_{m\to 0}\frac 1V \frac{\prod_k'(ij_{\nu=1,k}/(\Sigma V) +m)}
         {\prod_k (i j_{\nu=0,k}/(\Sigma V) +m)}
\\ &=&
\frac 1V \frac{\prod_k'ij_{\nu=1,k}\Delta / \pi }
         {\prod_k' i j_{\nu=1,k}^{as}\Delta / \pi }
\frac{\prod_k ij_{\nu=0,k}^{\rm as}\Delta / \pi }
         { \prod_k i j_{\nu=0,k}\Delta / \pi }
\frac{\prod_k' ij_{\nu=1,k}^{\rm as}\Delta /\pi }
 { \prod_k i j_{\nu=0,k}^{\rm as}\Delta /\pi  }.
\nn
\ee
The first two ratios are finite, but the third ratio has to be
regularized which we will do by a $\zeta$-function regularization
(see Appendix B). For $m =0$ we find the result (see (\ref{ratint-app})
\be
\frac 1V \frac{\prod_k'i  j_{\nu=1,k}^{\rm as}\Delta /\pi }
 {\prod_k i j_{\nu=0,k}^{\rm as} \Delta /\pi}
 =
\frac 1{V\Delta}
\frac{\Gamma^2(1-\delta)}{\Gamma^2(\frac 32 -\delta)}
=\frac{\Sigma^{(1)}}\pi
\frac{\Gamma^2(1-\delta)}{\Gamma^2(\frac 32 -\delta)}.
\label{ratint}
\ee
The ratios
${\prod_k' ij_{\nu,k} }/
         {\prod_k' i j_{\nu=1,k}^{as}}
$
can be easily evaluated numerically. We find
\be
\frac{\prod_k ij_{\nu=0,k} }
         {\prod_k i j_{\nu=0,k}^{as}} = \Gamma^2(\frac 34)/\sqrt 2, \qquad
\frac{\prod_k' ij_{\nu=1,k} }
         {\prod_k' i j_{\nu=1,k}^{as}}=\pi\Gamma^2(1.25)/2\sqrt 2.
\ee
Therefore,
\be
\Sigma^{(2)} = \Sigma^{(1)}.
\ee

Another way of evaluating the ratio (\ref{ratint}) is from the limit
\be
 \lim_{n \to \infty} \frac 1n
\frac{{\prod_{k=-n}^n}'i j_{\nu=1,k}^{\rm as}\Delta /\pi }
 {\prod_{k=-n}^n i j_{\nu=0,k}^{\rm as}\Delta /\pi }
=
 \lim_{n \to \infty} \frac 1n
\left [\frac{{\prod_{k=0}^{n-1}}(k +\frac 54)}
 {\prod_{k=0}^{n-1} (k+\frac 34) }\right]^{1/2}
\label{prod}
\ee
Using the infinite product representation of the $\Gamma$-function 
we find that this limit is given by
$
{\Gamma^2(\frac 34)}/{\Gamma^2(\frac 54)}.
$
We thus have
\be
\left .\frac{\prod_k'i  j_{\nu=1,k}^{\rm as}\Delta / \pi }
 {\prod_k i j_{\nu=0,k}^{\rm as} \Delta /\pi}\right |_\zeta
=\frac 1\Delta \lim_{n\to \infty}\frac 1{n}
\frac{{\prod_{k=0}^n}' i  j_{\nu=1,k}^{\rm as}\Delta /\pi }
 {\prod_{k=0}^n i  j_{\nu=0,k}^{\rm as}\Delta /\pi },
\ee
so that the definition (\ref{prod}) of the infinite product would have
resulted in an incorrect value for the chiral condensate. This explains
why in the random matrix calculation of previous section 
the incorrect chiral condensate is obtained from (\ref{zdemo}).
The fact that the two results differ by the eigenvalue at the cut-off is
in agreement with  
the interpretation that $\zeta$-function regularization 
includes the product of the eigenvalues up to a fixed energy.

\subsection{Random Gauge Field}.

In the condensed matter literature Dirac fermions in random gauge fields
have received a considerable amount of attention. Among the different types of
models that have been considered, 
we mention the random magnetic field problem \cite{Casher:1984tb},
the random flux model \cite{simons}, 
the random mass model \cite{ludwig} and the random gauge field model \cite{ludwig}. In this
section we only consider the latter model which is the quenched Schwinger
model with the gauge field
action is replaced by
\be
\frac 14 F_{\mu\nu}^2 \to \frac 12 \frac {g^2}{\sigma^2} A_\mu A_\mu.
\ee
The factor $g^2$ is introduced to have the same normalization as in 
\cite{ludwig}.
Also in this case we can make a Hodge decomposition of the gauge field
\cite{bernard},
\be
A_\mu = \epsilon_{\mu \nu} \del_\nu \phi + \del_\mu \Lambda.
\ee

Let us consider the action of an instanton configuration given by
\be
\phi = \frac 1{2g}\log (x^2+\rho^2).
\ee
In the Schwinger model the action of this configuration is given by
\be
S_{\rm inst} = \frac 12 \int d^2x  F_{01}^2 =\frac 12\int d^2x
\del^2 \phi \del^2 \phi= \frac \pi{3g^2\rho^2},
\ee
which is infrared finite. Also the anomalous contribution to the effective
action in the Schwinger model is infrared finite,
\be
-\int d^2x \frac{\mu^2}{2}\phi\partial^2\phi = 
\frac{1}{2}\big(\log \rho^2 +1 \big).
\ee
 For the random gauge model, on the other hand,  we obtain the
action 
\be
\frac 12 \frac{g^2}{\sigma^2}\int_{|x| < R} d^2x A_\mu^2 &=& 
\frac 12 \frac{g^2}{\sigma^2}\int_{|x| < R} 
d^2x \del_\mu\phi \del_\mu \phi\nn\\
&=&\frac{\pi}{2\sigma^2}\bigg[\log(\frac{R^2+\rho^2}{\rho^2})-
\frac{R^2}{R^2+\rho^2}\bigg].
\label{145}
\ee
which diverges for $R\to \infty$ which implies that instantons are
suppressed in the thermodynamic limit.  Therefore the chiral condensate
of the massless theory is zero, and by spectral duality, the eigenvalue
density of the Dirac operator vanishes independent of the topological
charge. For $\sigma^2 = \pi$ the suppression factor is $1/V$, which exactly
as in the Schwinger model for $N_f=1$, results in a finite spectral
density for $E\to 0$. Notice that the difference between the anomalous 
contribution in effective action of the Schwinger model and the random gauge 
action is only in the boundary term, which diverges for $R\to \infty$.

Let us evaluate the spectral density of 
a random gauge model with $N_f$ massless flavors 
and $n$ flavors 
with mass $m$. Because of the above remark, we restrict ourselves to 
the trivial topological charge sector, where the boundary 
term that gives the diverging result in (\ref{145}) is absent. 
Using that the pure gauge term decouples 
from the partition function, the effective action of this model 
is given by
\be
S_{\rm RG}=\int d^2x \Big[-\frac{1}{2}\phi
\lbrack\frac {g^2}{\sigma^2}\Delta +\mu^2(N_f+n)
\Delta\rbrack\phi
-i\frac{\theta g} {2\pi}\Delta\phi
+\sum_\alpha \bar{\psi}_\alpha\big\lbrack i\not \!\partial 
+me^{2g\gamma_5\phi}\big\rbrack\psi_\alpha\Big].
\ee
The massless fermion fields result in an overall constant which can be ignored. 
The action of the massive fermion fields will be replaced by their bosonized 
form given by
\be
S_{\rm F}(\phi,\varphi_\alpha) = \frac 12 \sum_{\alpha=1}^n (\del_\mu \varphi_\alpha)^2 -
cm \sum_{\alpha=1}^n \cos(2\varphi_\alpha \sqrt \pi -2ig\phi) .
\ee
If we define
\be
\Gamma(\phi) = -\frac 12\phi(\frac {g^2}{\sigma^2} \nabla^2 +\mu^2(N_f+n) \nabla^2) \phi
-i\frac{\theta g} {2\pi}\Delta\phi,
\ee
the total bosonic action is given by
\be
S_{\rm RG} = \int d^2x [\Gamma(\phi) + S_{F}(\phi,\varphi_\alpha)].
\ee 
After shifting the $\varphi_\alpha$ fields by
\be 
\varphi_\alpha \to \varphi_\alpha  +i\mu \phi -\frac \theta{2\sqrt \pi},
\ee
we obtain the action density
\be
\frac 12 \sum_{\alpha=0}^n (\del_\mu \varphi_\alpha)^2 +
\frac 12 (\frac{g^2}{\sigma^2}+\mu^2N_f)(\del_\mu\phi)^2
-
cm \sum_{\alpha=1}^n \cos(2\varphi_\alpha \sqrt \pi -\theta) 
-i\mu \del_\mu \phi \sum_{\alpha=1}^n \del_\mu \varphi_\alpha.\hspace{0.5cm}
\ee  
The final action density is obtained after performing the Gaussian
integral over $\phi$, 
\be
\frac 12\zeta 
\sum_{\alpha=1}^n \sum_{\beta=1}^n 
\del_\mu \varphi_\alpha
\del_\mu \varphi_\beta +\frac 12 \sum_{\alpha=1 }^n 
(\del_\mu \varphi_\alpha)^2 -
cm \sum_{\alpha=1}^n \cos(2\varphi_\alpha \sqrt \pi -\theta) ,
\label{bosact}
\ee
where
\be
\zeta = \frac{\mu^2}{\mu^2 N_f+g^2/\sigma^2}.
\ee
Considering the cosine term as a vertex, the propagator of the
$\varphi_\alpha$ fields is given by
\be
G_{\alpha \beta}(p) = \frac 1{p^2}
(\delta_{\alpha \beta} -\frac {\zeta}{1+n\zeta}).
\ee
Using this result one can easily derive the dependence of
the chiral condensate on the infrared cutoff which is taken to be equal to the
size of the box.
Following the derivation
of \cite{kenway} and using the lattice spacing $a$ as ultraviolet cutoff
we find for $a/L \ll 1$ that
\be
G_{\alpha\beta}(|x|\to 0) = -\frac 1{2\pi} (\delta_{\alpha\beta}
-\frac \zeta{1 +n \zeta})\log \frac aL
\ee
The chiral condensate is given by
\be
\Sigma &=& \frac 1n \frac 1V \del_m \log Z \sim 
\langle e^{2i\sqrt{\pi}\varphi_1} \rangle
\nn \\
&=&e^{-2\pi G_{11}(|x| \to 0)} \sim\left (\frac aL\right)^{1 -\zeta/(1+n\zeta)}
\ee
In the replica limit, $n \to 0$, we obtain the condensate
\be
\Sigma  \sim 
\frac 1{L^{1-\zeta}},
\ee
According to the Banks-Casher the smallest nonzero eigenvalue can be estimated
as
\be
\lambda_{\rm min} \sim \frac \pi{\Sigma V} \sim \frac 1{L^{1+\zeta }}.
\ee
On the other hand, if the eigenvalue density is given by
\be
\rho(\lambda) \sim V \lambda^\alpha,
\ee
the smallest nonzero eigenvalue follows from
\be
\int_0^{\lambda_{\rm min}} \rho(\lambda) d\lambda \approx 1.
\ee
Therefore,
\be
\lambda_{\rm min} \sim \frac 1{L^{2/(\alpha+1)}},
\ee
so that
\be
\alpha = \frac{1-\zeta}{1+\zeta}.
\ee
This result has also been derived from a more sophisticated
renormalization group analysis \cite{ludwig,fukui,carp}.
For the action (\ref{bosact}) we obtain
\be
\alpha =\frac{g^2+\mu^2\sigma^2(N_f-1)}{g^2+\mu^2\sigma^2(N_f+1)}
=\frac{1+\sigma^2(N_f-1)/\pi}{1+\sigma^2(N_f+1)/\pi}.
\label{specrf}
\ee  
In the quenched case ($N_f =0$), this result agrees with 
the result obtained in \cite{ludwig}, and in the limit
$\sigma \to \infty$ we recover the result for the $N_f$ flavor
Schwinger model \cite{smcon,smvac}. 
For $N_f=1$ the spectral density is qualitatively different
from the Schwinger model. A diverging spectral density for $N_f =0$ has
been observed in recent lattice simulations \cite{lat-poul}. 
If the spectral density vanishes, it still possible to derive sum rules
for the inverse Dirac eigenvalues 
\cite{poul-multi,janik,gernot-multi,janik-multi}. It could be that
such sum rules are easier obtained  for lattice simulation of
random Dirac fermions than 
for lattice simulations
of the Schwinger model with two massless flavors \cite{lat-poul}. 

A special case where the sum rules for the inverse Dirac eigenvalues
can be evaluated is $N_f =0$ and $n=1$. Since the sum rule follows
from the coefficient of $m^2$ in the expansion of the partition function
in powers of $m$, the spectral density corresponding to this sum rule
is the one of the Schwinger model with one massless flavor. 

For $n=1$ and $ N_f =0$ the partition function
after the transformation $\varphi_1 \to \mu\phi$
is given by
\be
Z(m,\theta)=\int D\varphi e^{\frac 12 F^2
\int d^2x \phi\Delta\phi +c m\frac 12\int d^2x [e^{-i\theta +2ig\phi(x)} + 
e^{i\theta- 2ig\phi(x)}]},
\nn \\
\label{finalrandom}
\ee  
where 
\be
F^2 =\mu^2+ \frac{\mu^2\sigma^2}\pi.
\ee
Let us evaluate the simplest sum rule in the sector of zero 
topological charge. To this end we expand the partition function
to second order in $m$ and integrate over $\theta$. We obtain
\be
\del_m^2 \log Z^{\nu=0}(m) = \int d^2x d^2y 
e^{4g^2 G^{\rm RG}(x,y)-4g^2G^{\rm RG}(0,0)},
\ee
where the random gauge field Green's function is given by
\be
G^{\rm RG}(x,y) =-\frac 1{4\pi F^2}\log(x-y)^2.
\ee
The volume dependence of the sum rule is thus given by
\be
\del_m^2 \log Z^{\nu=0}(m) \sim V^{2 -\mu^2/F^2}.
\ee
If the eigenvalue density scales as 
\be
\rho(E) \sim V E^{\alpha},
\ee
the volume dependence of the sum rule is given by
\be
\sum_k \frac 1{\lambda_k^2} \sim V^{\frac 2{1+\alpha}},
\ee
so that the exponent $\alpha$ is equal to
\be
\alpha = \frac{\mu^2}{2F^2 -\mu^2} = \frac 1{1+2\sigma^2/\pi}.
\ee
Indeed, this result agrees with (\ref{specrf}) for $N_f =1 $.

\section{Conclusions}
\label{conclusions}
In conclusion, 
we have analyzed Leutwyler-Smilga sum rules in the Schwinger model 
and found that they are in agreement with the universal result
from chiral perturbation theory and random matrix theory. Instead
of relying on general symmetry arguments, 
we have performed a microscopic calculation of the sum of the inverse
square Dirac eigenvalues in the sector of arbitrary topological charge
$\nu$, generalizing a result by Smilga obtained in the sector of
zero topological charge. 

We have shown that validity of the sum rules
follows from the clustering property of the scalar correlation functions.
This argument also applies to QCD with one flavor. For QCD with several 
flavors, the naive clustering argument fails due to the presence of 
Goldstone bosons. 
A modified clustering property is obtained from the chiral
Lagrangian that corresponds to 
the low energy limit of QCD partition function.

The dependence of the sum rules on the topological charge $\nu$  is
consistent with a shift of the Dirac eigenvalues by $\nu/2$ times the
average level spacing. Such shift of the Dirac spectrum exactly 
results in the chiral condensate for massless quarks in the sector
of topological charge $\nu=1$. However, obtaining this result, 
requires a regularization of the fermion determinant with a fixed energy 
cutoff such as for example the $\zeta$-function regularization.

In the microscopic derivation, the sum rules where obtained from more
general sum rules valid for a fixed external gauge field with topological
charge $\nu$. The most general sum rules in this class were conjectured
based a simplified derivation of Leutwyler-Smilga sum rules starting
from the
bosonized Schwinger model. It would be interesting to probe such
sum rules directly in lattice simulations.

The Schwinger model is part of a larger class of models known under the
name of random Dirac fermions which have been investigated in the 
context of the quantum Hall effect. The key difference between the two
models is that instantons are suppressed by the random gauge field action.
Therefore, the chiral condensate of the massless theory vanishes in the
domain below a critical value of the disorder.
In these models we expect
modified sum rules that  would be a useful probe for the scaling
behavior of the density of states near zero.

\vspace*{1cm}
\noindent{\bf Acknowledgements.}
We would like  to thank Andreas Ludwig, Andreas Smilga and 
Pierre van Baal for useful discussions. This work was supported in
part by US DOE Grant DE-FG-88ER40388.

\section{Appendix A}

In this appendix we calculate the integral given in (\ref{udiagint}).
We start from the result \cite{creutz},
\be
W(J)&=&\int_{SU(N_f)} dU \exp(Tr(JU))\nn \\
 &=&\sum_{i=0}^\infty 
\frac{2!\cdots (N_f-1)!}{i!\cdots (i+N_f-1)!}\Big(\det J\Big)^i.
\ee
We are interested in the integral
\be
\int_{SU(N_f)} dU  U_{11}^\nu U_{22}^\nu\cdots U_{N_f N_f}^\nu 
=\Bigg[\bigg(\frac{d}{dJ_{11}}\bigg)
^\nu\cdots\bigg(\frac{d}{dJ_{N_fN_f}}\bigg)^\nu W(J)\Bigg]\bigg\vert_{J=0}.
\ee
On the other hand, we can choose $J={\rm diag}[J_{11},J_{22},\cdots,
J_{N_fN_f}]$, so that 
\be
\det J = J_{11}J_{22}\cdots J_{N_fN_f}.
\ee
Using this result, 
we arrive at $N_f$ independent differentiations over 
diagonal matrix elements of $J$.
The only term in the sum that  contributes is the one with $i=\nu$
resulting in
\be
\int_{SU(N_f)} dU  U_{11}^\nu U_{22}^n\cdots U_{N_f N_f}^\nu 
&=& \Big[\nu!\Big]^{N_f}\frac{2!\cdots
  (N_f-1)!}{\nu!\cdots (\nu+N_f-1)!}\nn \\
&=&\Big[\nu!\Big]^{N_f}\prod_{i=0}^{N_f-1}\frac{i!}
{(i+\nu)!}.
\ee

\section{Appendix B}

In this appendix we compute the determinant of the Dirac operator
in $\zeta$-function regularization with eigenvalues given by the
asymptotic values of the zeros of Bessel functions,
\be
j_{\nu, k}^{\rm as} =  k+\frac \nu 2 -\delta \qquad {\rm with}\qquad
\delta = \frac 14.
\ee

In a $\zeta$ function regularization the determinant of the
Dirac operator in the sector of topological charge $\nu $ is given 
by
\be 
\log {\prod_k}' (i\Delta j_{\nu,k}^{\rm as}/\pi+m )
 = -
\left . \frac d{ds}\right |_{s=0}
\left [ \sum_{k=1}^\infty \frac 1{( k+\frac \nu 2 -\delta + im/\Delta)^s
\Delta^s }+
\sum_{k=1}^\infty \frac 1{( k+\frac \nu 2 -\delta -im/\Delta)^s
\Delta^s }\right ]. \nn \\
\ee
The function in between the brackets
is known as the Hurwitz zeta function,
\be
\zeta(s,a)\equiv \sum_{k=0}^\infty \frac 1{(k+a)^s}.
\ee
For convenience we only calculate the product for $m = 0$.
For $\nu = 1$ we find
\be
\log {\prod_k}' i\frac \Delta \pi j_{\nu=1,k}^{\rm as} 
&=& - 2 \left . \frac d{ds}\right |_{s=0}
[\zeta(s,\frac 32 -\delta) \Delta^{-s}] \nn\\
&=& -2\zeta'(0,\frac 32 -\delta) 
+2\zeta(0,\frac 32 -\delta ) \log\Delta.
\ee
Combining this with the result for $\nu = 0$, 
\be
\log  {\prod_k}' i\frac \Delta \pi j_{\nu=0,k}^{\rm as} &=& - 2 \left .
\frac d{ds}\right |_{s=0} 
[\zeta(s,1-\delta) \Delta^{-s}]\nn \\
&=& -2\zeta'(0,1 -\delta) +2\zeta(0, 1 -\delta) 
\log\Delta,
\ee
and using that
\be
\zeta(0,a) = \frac 12 -a, \qquad \zeta'(0,a) = \log\Gamma(
- \frac 12 \log(2\pi),
\ee
we find the ratio,
\be
\frac 1V \frac{\prod_k'i \frac \Delta \pi j_{\nu=1,k}^{\rm as} }
 {\prod_k i\frac \Delta \pi j_{\nu=0,k}^{\rm as} }
 =
\frac 1{V\Delta}
\frac{\Gamma^2(1-\delta)}{\Gamma^2(\frac 32 -\delta)}
=\frac{\Sigma^{(1)}}\pi
\frac{\Gamma^2(1-\delta)}{\Gamma^2(\frac 32 -\delta)}.
\label{ratint-app}
\ee

\end{document}